\documentclass[prl,twocolumn,showpacs,amsmath,amssymb,superscriptaddress]{revtex4-2}

\usepackage{amsmath,amssymb,amsfonts,bm,color,graphicx,tabularx}

\usepackage[unicode=true,colorlinks=true]{hyperref}

\usepackage[etex=true,export]{adjustbox}
\usepackage{makecell}
\usepackage{multirow}
\usepackage{amsmath}
\usepackage{verbatim}
\usepackage{longtable}

\hypersetup{linkcolor=blue, citecolor=blue,urlcolor=blue}
\usepackage{tabularx,multirow,array,diagbox}
\usepackage{adjustbox}
\usepackage{supertabular}

\newcommand{\beq}{\begin{equation}}
\newcommand{\eeq}{\end{equation}}
\newcommand{\beql}{\begin{equation*}}
\newcommand{\eeql}{\end{equation*}}
\newcommand{\beqn}{\begin{eqnarray}}
\newcommand{\eeqn}{\end{eqnarray}}

\begin{document}
\title{Spin Group Symmetry Criteria for Odd-parity Magnets}

\author{Xun-Jiang Luo}
%\email{luoxunjiang@ust.hk}
\affiliation{Department of Physics, Hong Kong University of Science and Technology, Clear Water Bay, 999077 Hong Kong, China}

\author{Jin-Xin Hu}
\affiliation{Department of Physics, Hong Kong University of Science and Technology, Clear Water Bay, 999077 Hong Kong, China}

\author{Mengli Hu}
\affiliation{Institute for Theoretical Solid State Physics, IFW Dresden, 01069 Dresden, Germany}

\author{K. T. Law}
\email{phlaw@ust.hk}
\affiliation{Department of Physics, Hong Kong University of Science and Technology, Clear Water Bay, 999077 Hong Kong, China}

\begin{abstract}

Odd-parity magnets (OPMs) have recently emerged as a new magnetic class, but their general symmetry criteria remain elusive. In this Letter, we establish these criteria through a comprehensive spin group symmetry analysis. Concretely, we identify eight distinct symmetry-driven cases that support OPMs with collinear, coplanar, or noncoplanar magnetic order. These are classified into three classes based on their spin textures: collinear (type-I), coplanar (type-II), and noncoplanar (type-III). From the Magndata database, we identify 33 candidate OPM materials and diagnose their spin-splitting character ($p$- or $f$-wave) by analyzing the representation of spin textures within an emergent Laue group derived from the spin space group, which reveals a variety of novel spin textures. To validate the symmetry criteria, we construct and analyze two theoretical models. Furthermore, we demonstrate that OPMs can host an intrinsic $\mathbb{Z}_2$ topology and propose a model for their realization. Our work provides a foundational framework for the future exploration of OPMs.

\end{abstract}
\maketitle
\textit{Introduction.}---
In recent years, unconventional magnetism has attracted widespread research interest in condensed matter physics \cite{chenhua2014,Nakatsuji2015,Binghai2017,Claudia2017,Šmejkal2018,Libor2020,Ma2021,Feng2022,ifmmode2022,Tomas2022,ifmmode2022a,BaiH2023,Krempaský2024,ZhouXiaodong2024,Reimers2024,BaiLing2024,McClarty2024,LeeSuyoung2024,Chen2025,HuMengli2025,Zhu2025}. Such magnetic systems exhibit symmetry-compensated magnetization and nonrelativistic spin splitting (NSS) \cite{Naka2019,Hayami2019,YuanLinDing2020}. This distinctive combination of properties makes them highly promising for spintronics applications \cite{Jakub2021,ZhangRun-Wu2024,LeiHan2024,Zhou2025,Zhang2025,hu2025nonlinear}. A prominent class is altermagnetism, which is characterized by collinear magnetic order and even-parity spin splitting \cite{ifmmode2022,ifmmode2022a}. Recently, odd-parity magnets (OPMs), such as $p$-wave systems, were proposed in coplanar magnetic orders, and have emerged  as a new research frontier \cite{HayamiSatoru2020, BirkHellenes2023,Brekke2024,zk69-k6b2,85fd-dmy8,YuPing2025,2025Minghuan,Huang2025,2025li,2025zhu,zhuang2025,2025liu}. OPMs exhibit spin splitting that reverses under momentum inversion, serving as a nonrelativistic analogue to Rashba-type spin–orbit coupling (SOC). In the original proposal, OPMs arise when spatial inversion is broken but the composite symmetry \(T\tau\) is preserved, where \(T\) denotes time reversal and \(\tau\) a fractional translation \cite{BirkHellenes2023}. Guided by this symmetry principle,  $p$-wave magnets have been experimentally realized \cite{Song2025a,Yamada2025a}. These advances raise a compelling question: can OPMs be realized in the absence of \(T\tau\) symmetry?

\begin{figure}
\centering
\includegraphics[width=3.3in]{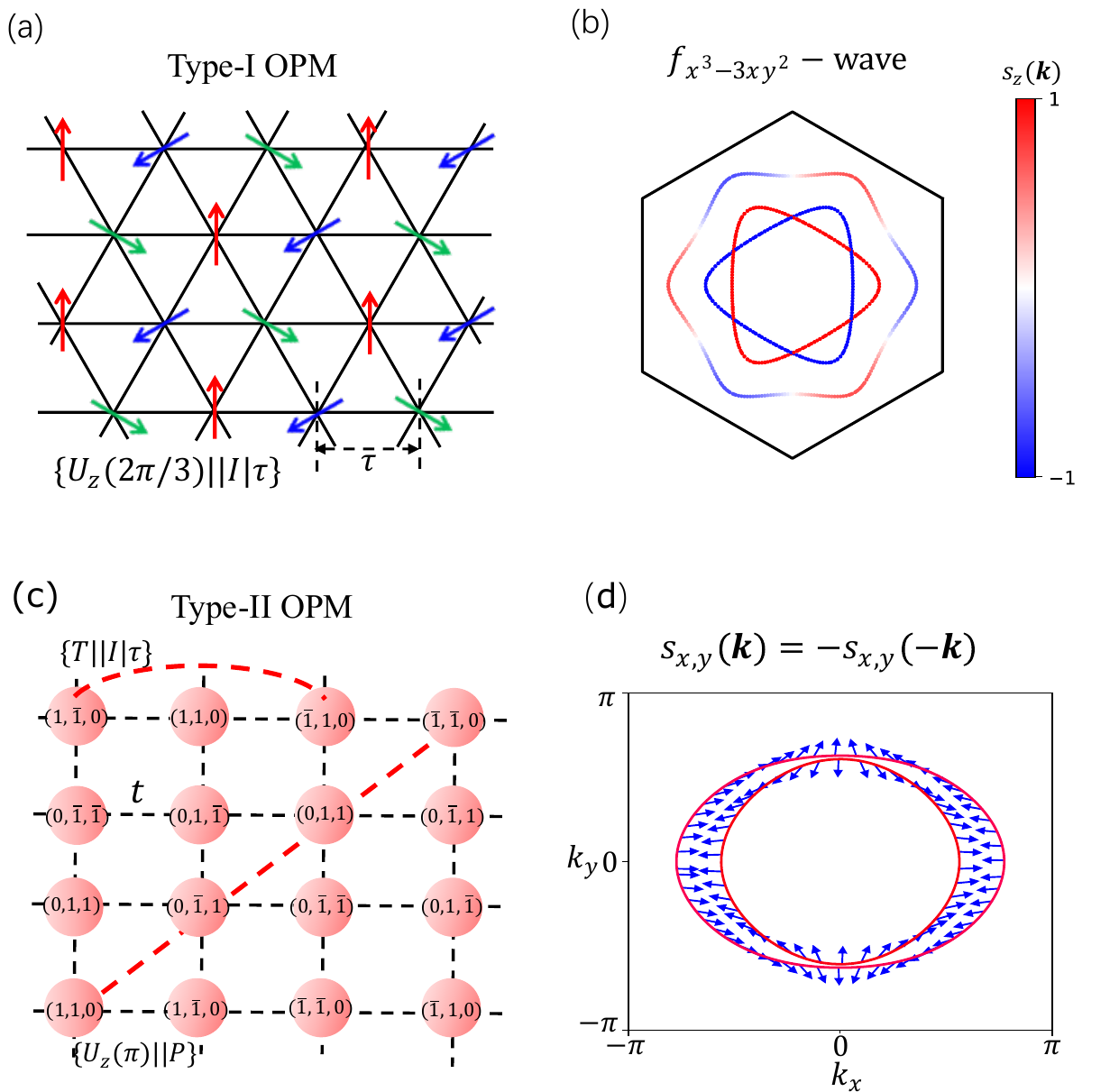}
\caption{ (a) Schematic illustration of $120^{\circ}$ antiferromagnetic order on a triangle lattice. (b) The isoenergy-surface characterized by $s_z(\bm k)$ for the model shown in (a). (c) Magnetic unit cell for a 2D lattice model on a  square lattice. The vectors $(x,y,z)$ denotes the magnetic moment directions. (d) The isoenergy-surface characterized by the vector $(s_x(\bm k,s_y(\bm k))$, shown by the blue arrow, for the lattice model in (c). We take $t=J=1$, $\mu=-2.5$ for (b),  and $\mu=-4.3$ for (d). }
\label{Fig1}
\end{figure}

The spin space group offers a comprehensive symmetry framework for systems with negligible SOC, where spin and spatial degrees of freedom are decoupled \cite{LiuPengfei2022, CheXiaobing2024, JiangYi2024, XiaoZhenyu2024,Xilin2025}. This framework has proven highly instrumental in classifying and understanding unconventional magnets \cite{Yuntian2025,2025song}. For example, the spin group description of altermagnets has guided their theoretical prediction \cite{ifmmode2022a}, materials design \cite{Mazin2023, Sheng2024, LiuYichen2024, SunWei2025}, and external control \cite{Sun2024a, GuMingqiang2025, CheYixuan2025, DuanXunkai2025}.
In contrast, the symmetry criteria governing OPMs have yet to be systematically established within the spin space group framework, despite preliminary proposals for their realization~\cite{BirkHellenes2023, YuPing2025,2025Minghuan,Huang2025,2025li,2025zhu,zhuang2025,2025liu}. Establishing such criteria is essential not only for a fundamental classification of OPMs, but also for enabling targeted material discovery, key steps toward bridging theory and application. Moreover, because of the existence of effective time-reversal symmetry, OPMs can also be topologically nontrivial, whereas the topological properties have yet to be investigated.

\begin{table*}
\centering
\setlength\tabcolsep{10pt}
\renewcommand{\arraystretch}{2}
\caption{Classification of OPMs and symmetry criteria for their emergence. Columns 3, 4, and 5 list the spin texture, symmetry criteria, and  magnetic orders for OPMs, respectively. Here, $A\bm k=\bm k$ and $B\bm k=-\bm k$, respectively. $C_2$ denotes the spin fliping operation in collinear magnetic orders. }
\begin{tabular}{|c|c|c|c|c|}
\hline
\multirow{9}{*}{\makecell{odd-\\ \\ parity\\ \\ magnets}}&types & spin textures & symmetry criteria & magnetic orders  \\
\cline{2-5}
 ~&\multirow{6}{*}{type-I} & \multirow{6}{*}{$s_z(\bm{k}) = -s_z(-\bm{k})$, $s_{x,y}(\bm{k}) = 0$} &  $\{C_2 ||B \}$ and $\nexists{\{C_2T||A\}}$   &  collinear   \\
\cline{4-5}
~&~&~& $\{T ||A | \tau\}$ and $\nexists{\{C_2T||A\}}$   &  collinear  \\
\cline{4-5}
~&~&~& (i) $\{U_z(\theta) ||A | \tau\}$ and $\{T ||A | \tau\}$   & noncoplanar \\
\cline{4-5}
~&~&~&  (ii) $\{U_z(\theta) ||A | \tau\}$ and $\{U_{x}(\pi)||B\}$    & \makecell{ 
 noncoplanar} \\
\cline{4-5}
~&~&~& (iii) $\{U_z(\theta) ||A | \tau\}$ and $\{TU_{z}(\pi)||A\}$  &  coplanar\\
\cline{4-5}
~&~&~& (iv)  $\{T||A|\tau\}$ and $\{TU_{z}(\pi)||A\}$& coplanar \\
\cline{4-5}
\cline{2-5}
~&type-II &$s_{x,y}(\bm k)=-s_{x,y}(-\bm k)$, $s_{z}(\bm k)=0$ &  $\{T||A|\tau\}$ and $\{U_{z}(\pi)||B\} $ &  noncoplanar \\
\cline{2-5}
 ~&type-III &$s_{x,y,z}(\bm{k}) = -s_{x,y,z}(-\bm{k})$&  $\{T||A|\tau\}$ & noncoplanar \\
\hline
\end{tabular}
\label{tab2}
\end{table*}

In this letter, we conduct a comprehensive spin group symmetry analysis  to establish general symmetry criteria for OPMs and show that they can be realized without $T\tau$ symmetry. By examining symmetry constraints on the spin textures of Bloch states, we identify eight distinct symmetry-driven cases for OPMs [Table~\ref{tab2}], which are classified into three types based on their spin textures: collinear (type-I), coplanar (type-II), and noncoplanar (type-III). From the Magndata database, we identify 33 candidate materials as OPMs [Table.~\ref{tab3}]. Remarkably, the spin textures in these systems form a representation of an emergent Laue group derived from the spin space group with odd parity. By deriving this representation,  we identify the spin-splitting character ($p$- or $f$-wave) in each material, which reveals a variety of novel spin textures. To validate the symmetry criteria, we construct two theoretical models and study their spin splitting properties. Finally, we propose and study a topological OPM in a bilayer breathing kagome lattice with nonplanar magnetic order.

\begin{table*}
\centering
\setlength\tabcolsep{3pt}
\renewcommand{\arraystretch}{2}
\caption{Candidate materials for type-I, type-II, and type-III OPMs. In type-I OPMs, the candidate materials in cases (i), (iii), (iv) satisfy the symmetry criteria associated with cases (i), (iii), (iv) in Table \ref{tab2}. Column 5 presents some typical SOC terms $\bm S(\bm k)\cdot \bm{\sigma}$ for describing NSS in candidate materials. }
\begin{tabular}{|c|c|c|c|c|}
\hline
\multirow{5}{*}{\makecell{OPMs}}&\multirow{3}{*}{\makecell{type-I}}&
(i) & Sr$_2$Fe$_3$Se$_2$O$_3$ &\multirow{4}{*}{\makecell{$k_i\sigma_i, (k_i+k_j)\sigma_i$\\
$k_y(k_y^2-3k_x^2)\sigma_z$,\\ $k_xk_yk_z\sigma_z$}}\\
\cline{3-4}
~&~& (iii)& \makecell{
CsFeCl$_3$, ThMn$_2$, EuIn$_2$As$_2$,
CsMnBr$_3$, RbFeCl$_3$,  RbNiCl$_3$,
Ba$_3$CoSb$_2$O$_9$}, CsFe(MoO$_4$)$_2$ & ~\\ 
\cline{3-4}
~&~& (iv) & \makecell{ Er$_2$Pt, DyBe$_{13}$, TbC$_2$, Ca$_2$Cr$_2$O$_5$,   Tm$_5$Pt$_2$In$_4$,  La$_{1/3}$Ca$_{2/3}$MnO$_3$\\ La$_{3/8}$Ca$_{5/8}$MnO$_3$,KFe(PO$_3$F)$_2$,
NiCr$_2$O$_4$, PrMn$_2$O$_5$,  GdMn$_2$O$_5$,
Sr$_2$FeO$_3$Cl,\\  
CoNb$_2$O$_6$, CeNiAsO,
DyMn$_2$O$_5$,BiMn$_2$O$_5$}& ~\\
\cline{2-5}
~&\multicolumn{2}{|c|}{type-II} & Ce$_3$InN & $k_x\sigma_x+k_y\sigma_y$\\
\cline{2-5}
~&\multicolumn{2}{|c|}{type-III} & \makecell{MgV$_2$O$_4$, Mn$_5$Si$_3$,   Ba(TiO)Cu$_4$(PO$_4$)$_4$,\\
Dy$_2$Co$_3$Al$_9$,  
DyFeWO$_6$,
Ho$_2$Cu$_2$O$_5$, BaFe$_2$Se$_3$} & \makecell{$\sum_i k_i\sigma_i, k_z\sum_i\sigma_i$}\\
\hline
\end{tabular}
\label{tab3}
\end{table*}

\textit{Symmetry criteria for OPMs}.---We begin by examining the spin group symmetry constraints on the momentum-space spin texture, defined as \(\bm{S}(\bm{k}) = (s_x(\bm{k}), s_y(\bm{k}), s_z(\bm{k}))\), where \(s_i(\bm{k}) = \langle \psi_{\bm{k}} | \sigma_i | \psi_{\bm{k}} \rangle\) for \(i = x, y, z\). Here, \(\sigma_{x,y,z}\) represent the Pauli matrices in the spin space, and \(|\psi_{\bm{k}}\rangle\) denotes the Bloch state at momentum \(\bm{k}\).
 To analyze these constraints, we introduce a spin-space group operation \(g = \{X_g U_g || R_g | \tau_g\}\), where \(\{R_g | \tau_g\}\) acts on the spatial coordinates (with \(R_g\) being a point group element and $\tau_g$ a fractional translation), and \(\{X_g U_g\}\) operates on the spin degrees of freedom. Here, \(X_g\) is either the identity operator \(I\) or the time-reversal operator \(T\), \(U_g \in \mathrm{SU}(2)\) is a spin rotation that can be represented by a \(\mathrm{SO}(3)\) operation. Under the action of \(g\), \(\bm{k}\) transforms as \(\bm{k} \to s_g R_g \bm{k}\), and the spin operator \(\bm{\sigma}\) transforms according to \(s_g {U}_g\), where \(s_g = +1\) (\(-1\)) corresponds to \(X_g = I\) (\(T\)).  Consequently, the vector \(\bm{S}(\bm{k})\) satisfies the symmetry constraint \cite{JiangYi2024,XiaoZhenyu2024}
\begin{equation}
\bm{S}(s_g R_g \bm{k}) = s_g {U}_g \bm{S}(\bm{k}).
\label{eq1}
\end{equation}
Since the fractional translation $\tau_g$ does not contribute to the constraints on \(\bm{S}(\bm{k})\), we take $\tau_g=0$ unless otherwise stated.

Notably, the definition of OPMs, namely $\bm{S}(\bm k)=-\bm{S}(-\bm k)$ with $\bm S(\bm k)\neq \bm {0}$, only involves $\bm{S}(\bm k)$ and $\bm{S}(-\bm k)$. Consequently, the symmetry criteria of OPMs can be fully determined by the symmetries that preserve and flip momentum. 
 The momentum-preserving symmetries, satisfying \( s_g R_g \bm{k} = \bm{k} \) and \( s_g U_g \neq I \), restrict some components of \(\bm{S}(\bm{k})\) to be zero. The momentum-flipping symmetries correspond to \( s_g R_g \bm{k} = -\bm{k} \) and \( s_g U_g \neq I \), which impose constraints between \(\bm{S}(\bm{k})\) and \(\bm{S}(-\bm{k})\), with excluding the inversion symmetry ($P$). The momentum-preserving symmetries include spin rotation combined with translation and space-time inversion with spin rotation, represented by:
\begin{align}
    g_1 &=\{ U_g || A | \tau_g\}, \quad g_2 =\{T U_g || B \},
    \label{eq3}
\end{align}
where  \( A \bm{k} = \bm{k} \) and \( B \bm{k} = -\bm{k} \), respectively. In three-dimensional (3D) systems, \( A = I \) and \( B = P \)  are standard choices. In 2D systems, possible selections include \( A = I \) or \( M_z \) (reflection in the \( z \)-plane), and \( B = P \) or \( R_{2z} \) (a \( \pi \) rotation around the \( z \)-axis).
The momentum-flipping symmetries include time-reversal combined with translation, space-inversion paired with spin rotation, and time-reversal coupled with spin rotation, expressed as:
\begin{equation}
g_3 = \{T || A | \tau_g\},\quad  g_4 = \{U_g || B \},\quad g_5 = \{T U_g || A \}.
\label{eq2}
\end{equation}
 In the following, we derive the symmetry criteria for OPMs according to the constraints generated by $g_{1-5}$.

The symmetry constraints imposed by \( g_{1-5} \) on $\bm S(\bm k)$ can be derived using the transformation rule in Eq.~\eqref{eq1} and they depend on the specific form of \( U_g \). To ensure compatibility with odd-parity NSS \cite{supp},  we choose \(U_g = U_{z}(\theta)\) (arbitrary rotation around the $z$-axis)  for \(g_1\), \(U_g = U_{z}(\pi)\)  for operations \(g_2\) and \(g_5\), and \(U_g = U_{x}(\pi)\) for \(g_4\). With these options, the symmetries $g_{1-5}$, respectively, generate the constraints
\beqn
&& \bm S(\bm k)=(0, 0,s_z(\bm k)),\nonumber\\
&& \bm S(\bm k)=(s_x(\bm k), s_y(\bm k),0),\nonumber\\
&& \bm S(\bm k)=(-s_x(-\bm k), -s_y(-\bm k),-s_z(-\bm k)),\nonumber\\
&& \bm S(\bm k)=(s_x(-\bm k), -s_y(-\bm k),-s_z(-\bm k)),\nonumber\\
&& \bm S(\bm k)=(s_x(-\bm k), s_y(-\bm k),-s_z(-\bm k)).
\label{eq4}
\eeqn
From these constraints, we observe that the symmetries \(g_1\) and \(g_2\) constrain two and one components of \(\bm {S}(\bm {k})\) to be zero, respectively. In contrast, the symmetries \(g_3\), \(g_4\), and \(g_5\) flip three, two, and one components of \(\bm {S}(\bm {k})\), respectively, under inversion of $\bm k$.  By combining these constraints, we classify OPMs into three types based on the dimensionality \(d_s\) of the symmetry-allowed span of \(\bm {S}(\bm {k})\),
\beqn
&&\text{type-I} \quad(d_s=1): \quad s_{i}(\bm{k}) = -s_i(-\bm{k}), s_{j,l}(\bm{k}) = 0;\nonumber\\
&&\text{type-II} \quad (d_s=2): \quad s_{i,j}(\bm{k}) = -s_{i,j}(-\bm{k}), s_{l}(\bm{k}) = 0;\nonumber\\
&&\text{type-III} \quad (d_s=3): \quad s_{i,j,l}(\bm{k}) = -s_{i,j,l}(-\bm{k}),
\eeqn
where \(i, j, l \in \{x, y, z\}\) and \(i \neq j \neq l\). Type-I OPMs, characterized by a collinear spin texture, can be realized through four kinds of combinations of symmetry operations: (i) \(g_1\) and \(g_3\); (ii) \(g_1\) and \(g_4\); (iii) \(g_1\) and \(g_5\),  and (iv) \(g_3\) and \(g_5\) \cite{supp}. In cases (i)-(iii), the $g_1$ symmetry constrains $\bm S(\bm k)$ along the $z$-direction, and $g_{3,4,5}$ symmetries flip the $z$-direction spin, leading to odd-parity NSS. Case (iv) corresponds to the initial proposal for the realization of $p$ -wave magnets \cite{BirkHellenes2023}, which belongs to case (iii) with the choice of $\theta=\pi$ in $g_1$ \cite{supp}. Type-II OPMs, exhibiting a coplanar spin texture, can be achieved by the combination of \(g_3\) and \(g_4\) symmetries. 
Type-III OPMs are realized in systems with noncoplanar magnetic orders under symmetry \(g_3\) alone.

In systems with collinear magnetic orders, spin is a good quantum number and $g_5$ symmetry is typically preserved. Since $g_5$ flips momentum $\bm k$ while preserving spin,    only even-parity NSS is allowed in such systems, corresponding to altermagnets \cite{ifmmode2022,ifmmode2022a}.  However, $g_5$ symmetry can be broken by the complex electron hopping terms \cite{YuPing2025,supp}. In such cases, OPMs can be realized in collinear antiferromagnetic systems where spin-opposite sublattices are related by inversion or translation, thereby preserving the symmetry $g_4$ or $g_3$. This scenario has recently been studied \cite{YuPing2025,2025Minghuan,Huang2025,2025li,2025zhu,zhuang2025,2025liu}. For coplanar magnetic orders, $g_5$ symmetry  remains intact, and type-I OPMs are realized when either $g_1$ or $g_3$ symmetry is additionally preserved.   The spin textures, symmetry criteria, and magnetic orders (collinear, coplanar, or noncoplanar) for OPMs, are summarized in columns 3-5 of Table~\ref{tab2}.

\textit{Candidate materials and NSS}.---
Using the online tool FINDSPINGROUP from Ref.~\cite{CheXiaobing2024} and the Bilbao Crystallographic Server, we identify 33 candidate materials as OPMs, as summarized in Table~\ref{tab3}. Among these candidates, 25, 1, and 7 materials belong to type-I, type-II, and type-III OPMs, respectively. In these systems, in addition to symmetries satisfying $g\bm{k} = \pm \bm{k}$, the symmetries $\tilde{g}$ associated with $\tilde{g}\bm{k} \ne \pm \bm{k}$ further impose constraints on $\bm{S}(\bm{k})$. The complete set of spin group symmetries ultimately determines the allowed partial-wave channels of NSS.

Under a spin space group \( G \), the spin texture \( \bm{S}(\bm{k}) \) forms a \( d_s \)-dimensional representation of the emergent point group \( \tilde{G} = \{s_g R_g \mid g \in G\} \) \cite{XiaoZhenyu2024}, which constrains the functional form of $\bm{S}(\bm{k})$. For the symmetry criteria in Table~\ref{tab2}, either $s_g=-1$ or $R_g=B$ (an inversion operation in momentum space) is satisfied. Consequently, the point group \( \tilde{G} \) for OPMs must include the inversion operation, thus belonging to a Laue group. Importantly, although the physical inversion symmetry of the crystal is broken, the emergent inversion symmetry in \( \tilde{G} \) ensures that \( \bm{S}(\bm{k}) \) exhibits odd parity.

For a given material, the symmetry-allowed partial-wave channels of NSS can be diagnosed by the
representation of $\bm S(\bm k)$ under the emergent Laue group $\tilde{G}$. For example, in type-I OPM CsFeCl$_3$, the spin space group is $ P^{2_{100}} 6_3 / ^1m^{2_{100}}m^1c|(3^1_{001}, 3^1_{001}, 1)^m 1$ and the emergent Laue group is $D_{6h}$. By Eq.~\eqref{eq1}, it can be shown that 
$s_z(\bm k)$ transforms as the $B_{2u}$ representation of $D_{6h}$, with $y(y^2-3x^2)$ as its lowest-order basis function. Near the $\Gamma=(0,0,0)$ point, this yields $s_z(\bm k)\approx k_y(k_y^2-3k_x^2)$, exhibiting a $f$-wave pattern. For clarity in presentation, we describe the resulting NSS in CsFeCl$_3$ using the effective term $k_y(k_y^2-3k_x^2)\sigma_z$. We emphasize that
this term captures only the sign structure of $s_z(\bm{k})$, not its detailed amplitude. Furthermore, 
 this term preserves the $T$ symmetry, explicitly broken in the actual system, and thus fails to capture $T$-breaking effects such as the lifting of Kramers degeneracy at time-reversal invariant momenta \cite{Kudasov2024}. In the Supplementary Material (SM) \cite{supp}, we systematically diagnose the NSS for each candidate material by identifying the representation formed by $\bm{S}(\bm{k})$, leading to a variety of effective terms  $\bm{S}(\bm{k})\cdot \bm{\sigma}$. A selection of these results is summarized in column 5 of Table~\ref{tab3}.

\textit{Theoretical models}.---
To validate the symmetry analysis, we construct two theoretical models, $H_1$ and $H_2$, representing type-I and type-II OPMs with coplanar [Fig.~\ref{Fig1}(a)] and noncoplanar [Fig.~\ref{Fig1}(c)] magnetic orders, respectively. The Hamiltonians $H_1$ and $H_2$ are expressed as:
\begin{equation}
H_{1,2} = \sum_{\langle ij\rangle,\sigma} t c_{i\sigma}^\dagger c_{j\sigma} 
+ J \sum_{i,\sigma,\sigma^{\prime}} \bm{m}_i \cdot c_{i\sigma}^\dagger \bm{\sigma}_{\sigma\sigma'} c_{i\sigma'},
\end{equation}
where $c_{i\sigma}^\dagger$ ($c_{i\sigma}$) denotes the creation (annihilation) operator for an electron with spin $\sigma = \uparrow, \downarrow$ at site $i$, and $\langle ij\rangle$ represents the summation over nearest-neighbor sites with hopping amplitude $t$. The second term describes the exchange coupling  with strength $J$.

The model $H_1$ describes a coplanar $120^\circ$ antiferromagnetic order on a triangular lattice with a $\sqrt{3} \times \sqrt{3}$ structure [Fig.~\ref{Fig1}(a)], comprising three magnetic sublattices (A, B, and C). The corresponding magnetic moments are $\bm{m}_\text{A} = (-\sqrt{3}/2, -1/2, 0)$, $\bm{m}_\text{B} = (\sqrt{3}/2, -1/2, 0)$, and $\bm{m}_\text{C} = (0, 1, 0)$. $H_2$ captures the key symmetries of the materials listed in the second row of Table~\ref{tab3}, including $g_1 = \{U_{z}(2\pi/3)|I|\tau\}$ and $g_5 = \{TU_z(\pi)||I\}$. Consequently, $H_1$ belongs to type-I OPMs with nonzero $s_z(\bm{k})$. We find that $s_z(\bm{k})$ transforms as a $B_1$ representation of $C_{6v}$, with a basis function $x(x^2 - 3y^2)$ \cite{supp}. Consequently, the NSS of $H_1$ can be described by $k_x(k_x^2 - 3k_y^2)\sigma_z$, which is consistent with an analytical derivation \cite{supp} and numerical verification [Fig.~\ref{Fig1}(b)]. Notably,  it can be shown that the effective SOC in $H_1$ is proportional to the hopping amplitude $t$ \cite{supp}, which is large in realistic systems.

The model $H_2$ is defined on a square lattice with sixteen magnetic sublattices. The arrangement of local magnetic moments within a unit cell is illustrated in Fig.~\ref{Fig1}(c). $H_2$ captures the key symmetries of CeIn$_3$N, including $g_3 = \{T||I|\tau\}$ and $g_4 = \{U_z(\pi)||P\}$, and therefore is a type-II OPM with nonzero $s_{x,y}(\bm{k})$. It can be shown that $\{s_x(\bm{k}), s_y(\bm{k})\}$ form a 2D representation $B_1 \oplus B_2$ of the point group $C_{2v}$ \cite{supp}, whose lowest order basis functions are $x\oplus y$. Consequently, the NSS of $H_2$ can be described by $k_x \sigma_x + k_y \sigma_y$ (the superposition coefficients are omitted), which is consistent with the numerical verification [Fig.~\ref{Fig1}(d)].

\begin{figure}
\centering
\includegraphics[width=3.3in]{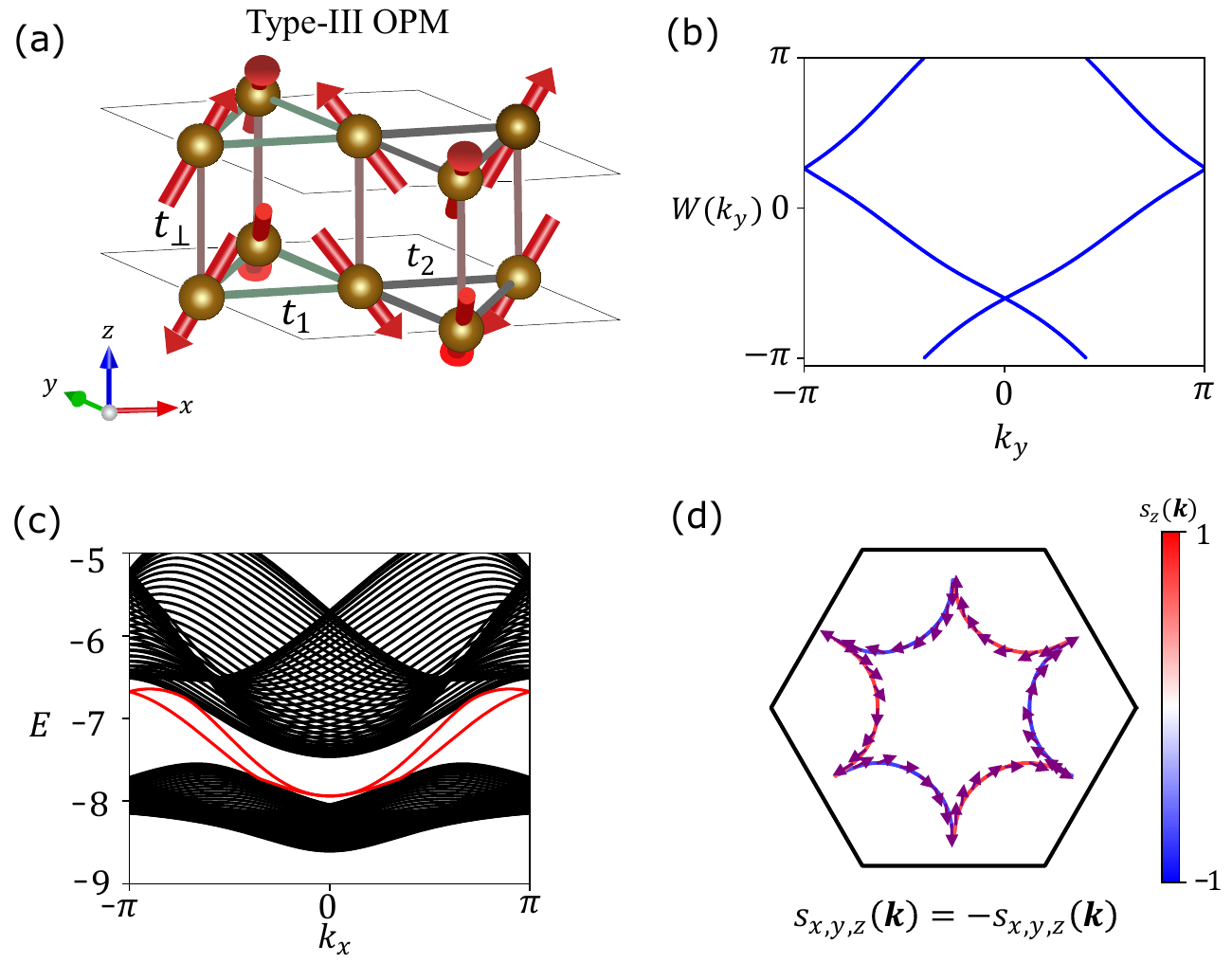}
\caption{(a) Schematic illustration of a bilayer kagome breathing lattice. (b) The wannier center $W(k_y)$ for the lowest energies two bands. (c) The energy spectrum for the system with a nanowire geometry along the $x$ direction. The red bands denote the edge states. (d) The spin polarization Fermi contour. The color encodes the value $s_z(\bm k)$. The pink arrows denote the direction of the vector $(s_x(\bm k), s_y(\bm k))$ at $\bm k$. See SM \cite{supp} for model parameters. }
\label{Fig3}
\end{figure}

\textit{Topological odd-parity magnets}.
OPMs can preserve an effective time-reversal symmetry $\tilde{T}$ satisfying $\tilde{T}^2 = -1$, which enables a nontrivial $\mathbb{Z}_2$  classification~\cite{Chiu2016}. To demonstrate such a topological phase, we construct a model on a bilayer breathing Kagome lattice [Fig.~\ref{Fig3}(a)]. This model features layer-contrasted magnetic moments with a finite canting angle, where the in-plane components form an all-in-all-out structure (see  SM \cite{supp} for details). For a single layer, the spin chirality induce Chern band topology and quantum spin Hall states are realized for the bilayer system,  protected by the symmetry $ \tilde{T}=\{T || M_z\}$ \cite{XiaoZhenyu2024}. Wilson loop calculations~\cite{YuRui2011} for the two lowest bulk bands confirm their nontrivial $\mathbb{Z}_2$ topology [Fig.~\ref{Fig3}(b)], and Fig.~\ref{Fig3}(c) shows the resulting helical edge states in a nanowire geometry. On the other hand, the inversion symmetry is broken in the breathing Kagome lattice. The whole system behaves as a type-III OPM due to the existence of $\tilde{T}$ symmetry.  Figure~\ref{Fig3}(d) depicts isoenergy Fermi surfaces characterized by odd-parity spin polarization. Analogously, it can be shown that $\bm S(\bm k)$ form a representation $E_1\bigoplus B_1$ of point group $C_{6v}$ and the NSS can be described by $k_y\sigma_x+k_x\sigma_y+k_x(k_x^2-3k_y^2)\sigma_z$ \cite{supp}.

\textit{Discussion and conclusion}.---We now discuss the the influence of SOC on OPMs. The odd-parity NSS is guaranteed by the existing spin group symmetries of $g_{1-5}$. However, when SOC is included, only the $T\tau$ symmetry ($g_3$) remains intact. Therefore, SOC  breaks the odd-parity spin splitting of systems without $T\tau$ symmetry.
 Furthermore,  the inclusion of SOC can significantly affect the topological properties of systems. For example, EuIn$_2$As$_2$ is a $p$-wave magnet \cite{k9p4tfhd} that can host surface Dirac cones when SOC effects are considered \cite{Riberolles2021}.

We emphasize that our framework also enables systematic exploration of even-parity NSS beyond altermagnets. For example, combining symmetries \(g_1\) and \(g_5\) with the choice of \(U_g = U_x(\pi)\) for \(g_5\) yields \(s_z(\bm{k}) = s_z(-\bm{k})\) while enforcing \(s_{x,y}(\bm{k}) = 0\). Furthermore, specific symmetries such as \(g_5\) alone in coplanar magnetic orders can produce hybrid-parity behaviors, where NSS exhibits even parity for \(s_{x,y}(\bm{k})\) and odd parity for \(s_z(\bm{k})\). We will systematically explore the symmetry criteria for these cases in future work, as such novel spin textures would bring spintronics applications.

OPMs are often effectively described using models that incorporate SOC and no $T$ symmetry breaking term is included \cite{Yamada2025a}. 
We emphasize that this description cannot accurately capture the band structures of OPMs, in which Kramers degeneracy at time-reversal invariant momenta can be generally lifted since $T$ symmetry is broken \cite{Kudasov2024}. This implies that the band structures are described by the interplay between effective SOC and Zeeman fields \cite{supp}, as best exemplified by 1D helimagnet (a minimum model of OPMs) \citep{Martin2012}. These unique electronic characteristics establish OPMs as a promising platform for investigating topological superconductivity \cite{Martin2012,2025sun}, which will be explored in our future work.

%OPMs are often modeled as systems exhibiting nonrelativistic SOC and the $T$ symmetry breaking effect is neglected. 

%and since the SOC terms preserve the $T$ symmetry, which is explicitly broken. The $T$ symmetry breaking can generally lift the Kramers degeneracy at time-reversal-invariant point \cite{Kudasov2024},  which 

In summary, we have established general symmetry criteria for OPMs through a comprehensive spin space group analysis. Based on these criteria, we have identified 33 candidate materials and systematically diagnosed their NSS forms according to the  formed representations by their spin textures. Our work  provides a solid foundation for future experimental and theoretical studies of OPMs.

\section{Acknowledgment}
We acknowledge useful discussions with Zhongyi Zhang, Junwei Liu, Xin Liu, Fengcheng Wu, and Xilin Feng.
K.T.L acknowledges the support from the Ministry of Science and Technology, China, and Hong Kong Research Grant Council through Grants No. 2020YFA0309600, No. RFS2021-6S03,
 No. C6025-19G, No. AoE/P-701/20, No. 16310520,
 No. 16307622, and No. 16309223.

\bibliography{reference}

\clearpage

\begin{widetext}
\begin{center}
\begin{large}
\textbf{Supplemental Material for ‘‘Spin Symmetries Criteria For Odd-parity Magnets"}
\end{large}
\end{center}

\setcounter{figure}{0}
\setcounter{equation}{0}
\renewcommand\thefigure{S\arabic{figure}}
\renewcommand\thetable{S\arabic{table}}
\renewcommand\theequation{S\arabic{equation}}

This Supplemental Material includes the following four sections:
(1) Symmetry criteria for odd-parity magnets (OPMs);
(2) OPMs in collinear magnetic orders;
(3) Theoretical models for OPMs;
(3) Candidate materials for OPMs.

\section{Symmetry conditions for odd-parity magnets}
In this section, we present the detailed derivation of symmetry criteria for OPMs. As discussed in the main text, five spin group symmetries $g_{1-5}$ fully determine the emergence of OPMs. These symmetries are:
\begin{align}
    g_1 &= \{ U_g \| A | \tau_g\}, \quad 
    g_2 = \{T U_g \| B \}, \quad 
    g_3 = \{T \| A | \tau_g\}, \quad  
    g_4 = \{U_g \| B \}, \quad 
    g_5 = \{T U_g \| A \},
    \label{eq3}
\end{align}
where \( A \bm{k} = \bm{k} \) and \( B \bm{k} = -\bm{k} \), respectively. Here, $U_g$ denotes a $\mathrm{SU}(2)$ rotation, which can be represented as a $\mathrm{SO}(3)$ matrix of the form $U_g = R_z(\alpha) R_y(\beta) R_x(\gamma)$, with the rotation matrices given by:
\begin{align}
R_z(\alpha) &= 
 \begin{bmatrix}
\cos \alpha & -\sin \alpha & 0 \\
\sin \alpha & \cos \alpha & 0 \\
0 & 0 & 1
\end{bmatrix}, \quad
R_y(\beta) =
\begin{bmatrix}
\cos \beta & 0 & -\sin \beta \\
0 & 1 & 0 \\
\sin \beta & 0 & \cos \beta
\end{bmatrix}, \quad
R_x(\gamma) =
\begin{bmatrix}
1 & 0 & 0 \\
0 & \cos \gamma & -\sin \gamma \\
0 & \sin \gamma & \cos \gamma
\end{bmatrix},
\label{seq2}
\end{align}
where $\alpha$, $\beta$, and $\gamma$ represent rotation angles about the $z$-, $y$-, and $x$-axes, respectively. The symmetries $g_{1-5}$ impose constraints on the spin texture $\bm{S}(\bm{k})$ through the relation:
\begin{equation}
\bm{S}(s_g R_g \bm{k}) = s_g {U}_g \bm{S}(\bm{k}),
\label{seq1}
\end{equation}
where $\bm{S}(\bm{k}) = (s_x(\bm{k}), s_y(\bm{k}), s_z(\bm{k}))$ denotes the spin texture vector, with components $s_i(\bm{k}) = \langle \psi_{\bm{k}} | \sigma_i | \psi_{\bm{k}} \rangle$ for $i = x, y, z$. Here, $\sigma_i$ represents the Pauli matrix in the spin space, $|\psi_{\bm{k}}\rangle$ is the Bloch state at momentum $\bm{k}$, and the factor $s_g = \pm 1$ indicates the absence ($+1$) or presence ($-1$) of time-reversal operation in the symmetry.

The constraint in Eq.~\eqref{seq1} depends on the specific form of $U_g$. For symmetry $g_1$, which leaves $\bm{k}$ invariant, a nonzero rotation around two axes in $U_g$ enforces $\bm{S}(\bm{k}) = \bm{0}$ according to Eqs.~\eqref{seq2} and \eqref{seq1}. For symmetry $g_2$, it can be shown that $\bm{S}(\bm{k})$ remains nonzero only when $U_g$ corresponds to a $\pi$ rotation about a single axis. Therefore, without loss of generality, we adopt $U_g = U_z(\theta)$ (arbitrary rotation about the $z$-axis) for $g_1$ and $U_g = U_z(\pi)$ for $g_2$. For symmetries $g_4$ and $g_5$, which reverse momentum $\bm{k}$, arbitrary rotation angles $\theta$ and $\phi$ in $U_g$ are formally allowed. However, considering the compatibility with odd-parity NSS which requires at least one component of $\bm{S}(\bm{k})$ to change sign under momentum inversion, only $\pi$ rotation and rotation around a single axis is allowed for $g_4$ and $g_5$, respectively. Consequently, we choose $U_g = U_x(\pi)$ for $g_4$, and without loss of generality, $U_g = U_z(\pi)$ for $g_5$. With these specific choices, the symmetries $g_{1-5}$ generate the following constraints:
\beqn
&& \bm S(\bm k)=(0, 0,s_z(\bm k)),\nonumber\\
&& \bm S(\bm k)=(s_x(\bm k), s_y(\bm k),0),\nonumber\\
&& \bm S(\bm k)=(-s_x(-\bm k), -s_y(-\bm k),-s_z(-\bm k)),\nonumber\\
&& \bm S(\bm k)=(s_x(-\bm k), -s_y(-\bm k),-s_z(-\bm k)),\nonumber\\
&& \bm S(\bm k)=(s_x(-\bm k), s_y(-\bm k),-s_z(-\bm k)).
\label{eq4}
\eeqn
Under these constraints, we can classify OPMs into three types
based on the dimensionality \(d_s\) of the symmetry-allowed span of \(\bm {S}(\bm {k})\) in momentum space:
\beqn
&&\text{type-I} \quad(d_s=1): \quad s_{i}(\bm{k}) = -s_i(-\bm{k}), s_{j,l}(\bm{k}) = 0;\nonumber\\
&&\text{type-II} \quad (d_s=2): \quad s_{i,j}(\bm{k}) = -s_{i,j}(-\bm{k}), s_{l}(\bm{k}) = 0;\nonumber\\
&&\text{type-III} \quad (d_s=3): \quad s_{i,j,l}(\bm{k}) = -s_{i,j,l}(-\bm{k}),
\eeqn
where \(i, j, l \in \{x, y, z\}\) and \(i \neq j \neq l\). 

For type-I OPMs with collinear spin texture, the symmetry \(g_1\) is useful as it constrains $s_{x,y}(\bm k)=0$. By combining $g_1$ symmetry and other symmetries, there are three cases for the realization of type-I OPM:  (i) $g_1$ and $g_3$, (ii) $g_1$ and $g_4$, (iii) $g_1$ and $g_5$. In addition, type-I OPMs can also be realized under the following symmetry combinations: (iv) $g_3$ and $g_5$, (v) $g_2$, $g_4$, and $g_5$ with an alternative choice $U_g=U_x(\pi)$ in $g_2$. 
However, both  cases (iv) and (v) are not independent of cases (i)-(iii). For case (iv), the presence of symmetries $g_3=\{\mathcal{T}||A|\tau\}$ and $g_5=\{TU_z(\pi)||A\}$ implies the existence of symmetry $g_1=g_3g_5=\{U_z(\pi)||I|\tau\}$. Therefore, case (v) belongs to case (iii). For case (v), the presence of symmetries $g_2=\{TU_x(\pi)||P\}$ and $g_4=\{U_x(\pi)||P\}$ indicates the existence of symmetry $T=g_2g_4$. However, since the $T$ symmetry is explicitly broken, a fractional translation must be present in \(g_2\) or \(g_4\). Without loss of generality, we assign the fractional translation to $g_4=\{U_x(\pi)||P|\tau\}$. In this case, the symmetry $g_1=g_2g_4g_5=\{U_z(\pi)||A|\tau\}$ is preserved. Therefore,  case (v) is not independent and belongs to cases (i),   (ii), and (iii).

For type-II OPMs with coplanar spin textures, the symmetry $g_2$ is useful, which constrains $s_z(\bm k)=0$.
By combining $g_2$ symmetry and other symmetries, there are two cases for the realization of type-II OPMs:  (a) $g_2$ and $g_3$, (b) $g_2$ and $g_4$ with an alternative choice $U_g=U_z(\pi)$ in $g_4$. However, the presence of symmetries $g_2=\{TU_z(\pi)||B\}$ and $g_4=\{U_z(\pi)||B\}$ implies the existence of symmetry $T=g_2g_4$. This implies that there is a fractional translation $\tau$ in symmetry $g_2$ or $g_4$. Without loss of generality, we take $g_4=\{U_z(\pi)||B|\tau\}$. In this case, the symmetry $g_3=g_2g_4$ is preserved, indicating that case (b) reduces to case (a). In addition, type-II OPMs can also be realized under the combination of $g_3$ and $g_4$ symmetries, denoted as case (c). However,  case (c) is also not independent. In this case, the symmetry $g_2=g_3g_4$ can be obtained. For type-III OPMs, the symmetries \(g_1\) and \(g_2\) are forbidden, as they impose constraints on the dimensionality of \(\bm{S}(\bm{k})\). Type-III OPMs can be realized under the symmetry \(g_3\) alone.

\begin{figure}
\centering
\includegraphics[width=5.8in]{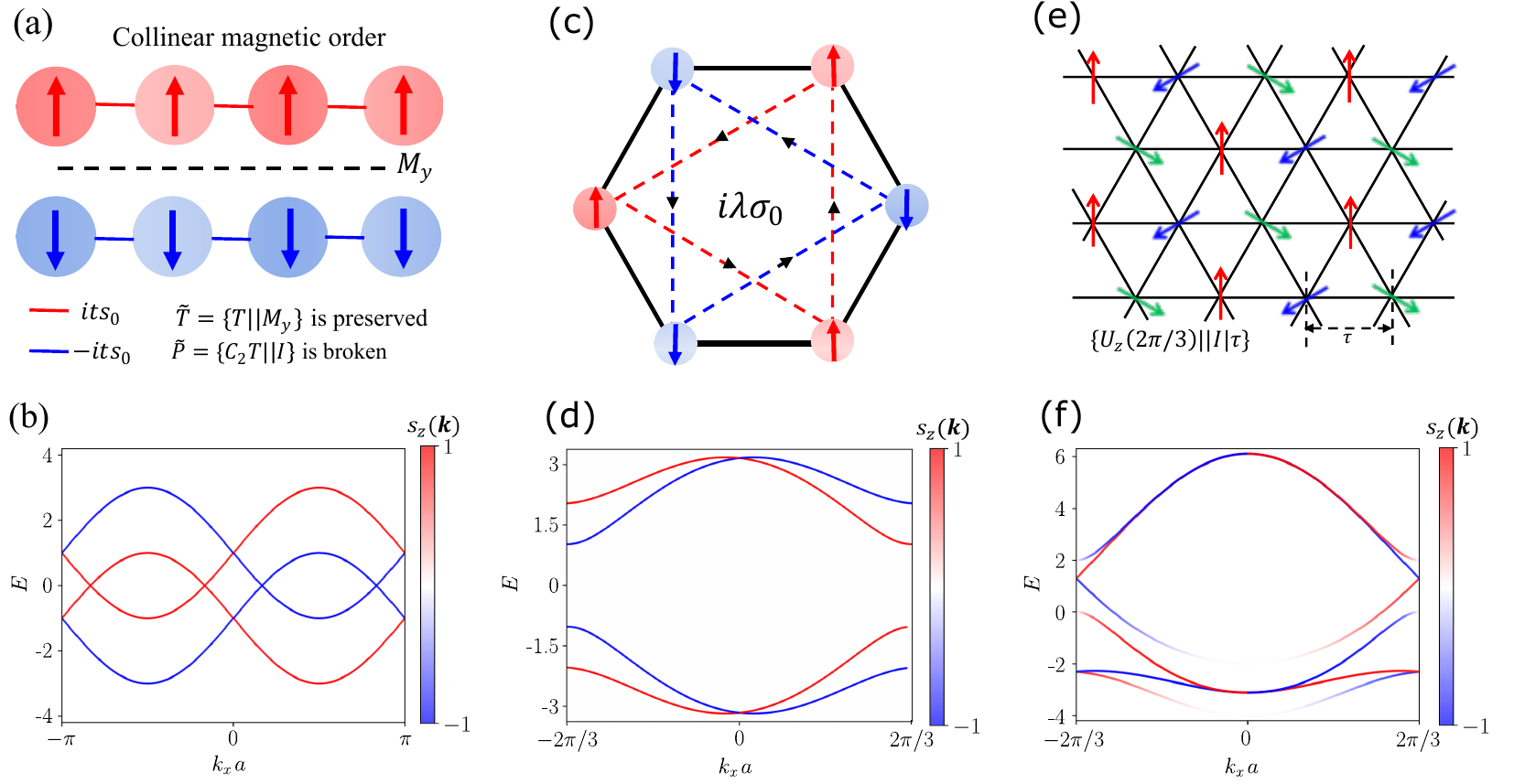}
\caption{ (a) Schematic illustration of a 1D  OPM. (b) The energy bands for the illustrated model in (a). (c) Schematic illustration of Haldane model with antiferromagnet order. (d) The energy bands of the illustrated model in (c). (e) 
Schematic illustration of $120^{\circ}$ antiferromagnet order on a triangular lattice. (f) The energy bands of the illustrated model in (f). 
 We take $t=J=1$ and $\lambda=0.2$ in (d). }
\label{Fig4}
\end{figure}

\section{OPMs in collinear magnetic orders}
\label{AppendixB}
In this section, we demonstrate that OPMs can be realized in systems with collinear magnetic orders and present two theoretical models to illustrate this. In systems with collinear magnetic orders, early studies suggest that only even-parity nonrelativistic spin splitting (NSS) is allowed, leading to altermagnets~\cite{ifmmode2022, ifmmode2022a}. This is attributed to the effective inversion symmetry $C_2T$ (with $C_2$ being a spin flip operation) for collinear magnetic orders, which flips the momentum \(\bm{k}\) while preserving the spin. However, this argument overlooks the electron hoppings, which can break the $C_2T$ symmetry. In the following, we use two theoretical models to demonstrate how OPMs can emerge in such systems.

As illustrated in Fig.~\ref{Fig4}(a), the system consists of two sublattices: sublattice A (red atoms) and sublattice B (blue atoms), which have opposite spin orientations and are related by a mirror operation \(M_y\). We consider the complex hopping for sublattices A and B, which are \(it\) and \(-it\), respectively, and therefore the \(C_2 T\) symmetry is broken. The model Hamiltonian for this one-dimensional system is given by:
\begin{align}
\mathcal{H}_1 = J \tau_z \sigma_z + 2t \sin( k_xa) \tau_z \sigma_0,
\end{align}
where the first term represents the onsite exchange interaction with strength \(J\),  the second term describes the sublattice-dependent complex hopping with magnitude \(t\), and $a$ is lattice constant. The Pauli matrices \(\tau_z\) acts on the orbital degrees of freedom and \(\sigma_0\) is the identity matrix in the spin space. $\mathcal{H}_1$ respects an effective time-reversal symmetry \(\tilde{T} = \tau_x \sigma_y K\), where \(K\) denotes complex conjugation, but breaks the symmetry \(C_2 T = \sigma_z K\). According to the symmetry criteria in Table~\ref{tab2}, \(\mathcal{H}_1\) describes a type-I OPM.
The energy spectrum of $\mathcal{H}_1$ is $E_s=\pm J+2ts\sin k_x$, where $s=1$ ($-1$) for spin up (down). Therefore, the NSS is $E_{+}-E_{-}=4t\sin k_x$, exhibiting a $p_x$-wave pattern. The energy bands of $\mathcal{H}_1$ is plotted in Fig.~\ref{Fig4}(b).

The Haldane model with collinear antiferromagnet order, as illustrated in Fig.~\ref{Fig4}(c), is a typical example to realize OPMs and was firstly pointed out in Ref.~\onlinecite{YuPing2025}. The model Hamiltonian can be written as 
\beqn
&&\mathcal{H}_2(\bm k)=f_x(\bm k)\tau_x\sigma_0+f_y(\bm{k})\tau_y\sigma_0+f_z(\bm{k})\tau_z\sigma_0+J\tau_z\sigma_z,\nonumber\\
&& f_x(\bm k)=t(1+2\cos (\sqrt{3}k_xa/2)\cos(3k_ya/2)),\nonumber\\
&&f_y(\bm k)=t\cos (\sqrt{3}k_xa/2)\sin(3k_ya/2), \nonumber\\
&&f_z(\bm k)=\lambda(2\sin (\sqrt{3}k_xa)-4\sin (\sqrt{3}k_xa/2)\cos(3k_ya/2)).
\eeqn
where $\lambda$ is the next-nearest-neighbor hopping amplitude. $\mathcal{H}_2$ respects the effective time-reversal symmetry $\tilde{T}=\tau_x\sigma_yK$ and the $C_2T$ symmetry is broken.  Therefore, $\mathcal{H}_2$ describes a type-I OPM according to the symmetry criteria in Table \ref{tab2}. Figure.~\ref{Fig4}(d) plots the energy bands of $\mathcal{H}_2$.

The distribution pattern of $s_z(\bm{k})$ in the momentum space  can be directly revealed from the energy spectrum of $\mathcal{H}_2$, which is 
\beqn
E_{r,s}(\bm k)=r\sqrt{f_x^2(\bm k)+f_y^2(\bm k)+(f_z(\bm k)+sJ)^2},
\eeqn
with $s,r=\pm 1$ with $s=1$ ($s=-1$) for spin up (down).
The energy of the first band of $\mathcal{H}_2$ is $E_1=\sqrt{f_x^2(\bm k)+f_y^2(\bm k)+(|f_z(\bm k)|+J)^2}$, which is associated with $s=1$ ($s=-1$) when $f_z(\bm k)>0$ ($f_z(\bm k)<0$). Therefore, $s_z(\bm k)$ for the Bloch states associated with the first energy band has the same sign structure of $f_z(\bm k)$. By expanding $f_z(\bm k)$ around $\Gamma=(0,0)$ point to the third-order of $\bm k$, we have $f_z(\bm k)\approx -3\sqrt{3}\lambda a^3/4k_x(k_x^2-3k_y^2)$, where we have used the relations $\sin x\approx x-x^3/6$ and $\cos x\approx 1-x^2/2$. Therefore, $s_z(\bm k)\propto  k_x(k_x^2-3k_y^2)$, which exhibits a $f$-wave pattern. 

%Moreover, the first and second bands have the opposite spin and the energy gap between them is $E_{1,1}-E_{1,-1}$, which is zero when $f_z(\bm k)=0$

\section{Theoretical models for OPMs}
In this section, we present the details for the introduced three theoretical models in the main text, which realize type-I, type-II, and type-III OPMs, respectively.

\subsection{Theoretical model for type-I OPMs}
We consider the 120° antiferromagnetic order on a triangular lattice [see Fig.~\ref{Fig4}(e)] with a $\sqrt{3} \times \sqrt{3}$ reconstruction. The unit cell comprises three sublattices, denoted as A, B, and C. The local magnetic moments on these sublattices are given by $\bm{m}_\text{A} = (-\sqrt{3}/2, -1/2, 0)$, $\bm{m}_\text{B} = (\sqrt{3}/2, -1/2, 0)$, and $\bm{m}_\text{C} = (0, 1, 0)$.
The model Hamiltonian in real space is expressed as:
\beqn
H_1(\bm r)=\sum\limits_{\bm{r},\sigma,\sigma^{\prime}}J\bm{m}(\bm r)\cdot c_{\sigma}^{\dagger}(\bm r) \bm{\sigma} c_{\sigma^{\prime}}(\bm r)+t\sum_{\langle \bm{r}\bm{r}^{\prime}\rangle,\sigma}c_{\sigma}^{\dagger}(\bm r)c_{\sigma}(\bm r^{\prime}),
\eeqn
where $\bm{m}(\bm{r}) = \left( \cos(\bm{K} \cdot \bm{r} + \phi), \sin(\bm{K} \cdot \bm{r} + \phi), 0 \right)$ represents the spatially varying magnetization, $\bm{\sigma} = (\sigma_x, \sigma_y, \sigma_z)$ is the vector of Pauli matrices, $\bm{K} = \frac{4\pi}{3a}(1/2, \sqrt{3}/2)$ is the propagation-vector of antiferromagnet order, and $\phi = -5\pi/6$ is the phase offset. Here,  $\langle \bm{r} \bm{r}^{\prime} \rangle$ denotes the summation over nearest-neighbor sites.
The lattice sites belonging to the A, B, and C sublattices are located at
\begin{align*}
\bm{r}_A = m \bm{e}_1 + n \bm{e}_2, \quad
\bm{r}_B = m \bm{e}_1 + n \bm{e}_2 + \bm{a}_1, \quad
\bm{r}_C = m \bm{e}_1 + n \bm{e}_2 + \bm{a}_2,
\end{align*}
respectively, where $\bm{e}_1 = \left(3/2, \sqrt{3}/2\right)a$ and $\bm{e}_2 = \left(3/2, -\sqrt{3}/2\right)a$ are the basis vectors of the magnetic supercell, and $\bm{a}_1 = (1, 0)a$ and $\bm{a}_2 = \left(1/2, \sqrt{3}/2\right)a$ are the primitive vectors of the underlying triangular lattice. Accordingly, the local moments take the values $m(\bm r)=\bm{m}_{\text{A}}$, $m(\bm r)=\bm{m}_{\text{B}} $, and $m(\bm r)=\bm{m}_{\text{C}}$ for lattice sites belong to the A, B, and C sublattices, respectively.

After Fourier transformation, the Bloch Hamiltonian in the sublattice and spin space is given by:
\beqn
H_1(\bm k)=\begin{pmatrix}
J(-\sqrt{3}/2\sigma_x-\sigma_y/2) & T_{\text{AB}}(\bm k)\sigma_0 & T_{\text{AC}}(\bm k) \sigma_0  \\
T_{\text{BA}}(\bm k)\sigma_0 & J(\sqrt{3}/2\sigma_x-\sigma_y/2) & T_{\text{BC}}(\bm k)\sigma_0\\
T_{\text{CA}}(\bm k) \sigma_0& T_{\text{CB}}(\bm k)\sigma_0 & J\sigma_y\\
\end{pmatrix},
\eeqn
where $T_{\alpha\beta}(\bm{k}) = T_{\beta\alpha}^*(\bm{k})$, $T_{\text{AB}}(\bm{k}) = T_{\text{BC}}(\bm{k}) =T_{\text{AC}}^{*}(\bm{k})= t\left[ e^{i k_x a} + e^{i(-k_x a/2 + \sqrt{3}k_y a/2)} + e^{-i(k_x a/2 + \sqrt{3}k_y a/2)} \right]$. 
By diagonalizing the Bloch Hamiltonian $H_1(\bm{k}) |\psi_{n\bm{k}}\rangle = E_n(\bm{k}) |\psi_{n\bm{k}}\rangle$, the energy bands $E(\bm{k})$ can be obtained, as shown in Fig.~\ref{Fig4}(f). 

%The spin expectation value $s_{x,y}(\bm{k})=0$ as the existence of the $g_1$ symmetry and $s_z(\bm k)=-s_z(\bm k)$ as the $g_5$ symmetry.

\begin{comment}
$H_3(\bm k)$ respects the symmetry $g_1=Fe^{-i2\pi/3\sigma_z}$ and $g_5=\sigma_xK$, with 
\beqn
F=\begin{pmatrix}
0& 1& 0\\
0&0&1\\
1&0&0
\end{pmatrix}.
\eeqn
\end{comment}

Since $H_1$ respects the symmetries $g_1=\{U_z(2\pi/3)||I|\tau\}$ and $g_5=\{TU_z(\pi)||I\}$, it belongs to a type-I OPM where only the spin expectation value $s_z(\bm k)$ is nonzero. In addition to $g_1$ and $g_5$, $H_1$ also preserves the symmetries $\hat{C}_{3z}=\{I||R_{3z}\}$, $\hat{M}_y=\{I||M_y\}$, and $\hat{C}_{6z}=g_5\hat{C}_{3z}=\{TU_z(\pi)||R_{3z}\}$. The symmetries $\hat{C}_{6z}$ and $\hat{M}_y$ transform the momentum as $\hat{C}_{6z}\bm k=R_{6z}\bm k$ and $\hat{M}_y\bm k=M_y\bm k$, respectively. These two symmetries serve as generators of the point group $\tilde{G}=C_{6v}$. According to Eq.~\eqref{seq1}, we have  the symmetry constraints
\beqn
s_z(\bm k)=-s_z(\hat{C}_{6z}\bm k), \quad s_z(\bm k)=s_z(\hat{M}_y\bm k).
\eeqn
Therefore, the spin texture $s_z(\bm k)$ forms a $B_1$ representation of $C_{6v}$, the lowest-order basis function of which is $x(x^2-3y^2)$. Consequently, near the $\Gamma=(0,0)$ point, we have $s_z(\bm k)\approx k_x(k_x^2-3k_y^2)$. In the following, we provide an analytical derivation of this result.

%Therefore,  these spin group symmetries lead to the point group $\tilde{G}=C_{6v}$. According to the Eq.~\ref{seq1}, it can be shown that $and the spin textures 

%To reveal the distribution pattern of spin expectation value $s_z(\bm k)$ in the momentum space, 
We apply a local basis transformation to the real-space Hamiltonian  $H_1(\bm r)$
\beqn
\tilde{H}_1(\bm r)=\mathcal{U}(\bm r)H_1(\bm r)\mathcal{U}^{\dagger}(\bm r),\quad \mathcal{U}(\bm r)=e^{i\bm{K}\cdot \bm{r}\sigma_z/2}.
\eeqn
Notably, the transformation satisfies  \( \mathcal{U}(\bm{r}) \bm{m}(\bm{r}) \cdot \boldsymbol{\sigma} \mathcal{U}^\dagger(\bm{r}) = \sigma_x \), indicating that \( \mathcal{U}(\bm{r}) \) rotates all local magnetic moments to align along the \( x \)-direction.  The nearest-neighbor hopping term transforms as follows
\beqn
\sum_{\langle \bm{r}\bm{r}^{\prime}\rangle,\sigma}c_{\sigma}^{\dagger}(\bm r)c_{\sigma}(\bm r^{\prime})&=&\sum_{\bm{r},\sigma}[c_{\sigma}^{\dagger}(\bm r)c_{\sigma}(\bm r+\bm{a}_1)+c_{\sigma}^{\dagger}(\bm r)c_{\sigma}(\bm r-\bm{a}_2)+c_{\sigma}^{\dagger}(\bm r)c_{\sigma}(\bm r+\bm{a}_3)+h.c.]\nonumber\\
&=&\sum_{\bm{r},\sigma}[e^{-i\bm{K}\cdot \bm{a_1}[\sigma_z]_{\sigma\sigma}/2}\tilde{c}_{\sigma}^{\dagger}(\bm r)\tilde{c}_{\sigma}(\bm r+\bm{a}_1)+e^{i\bm{K}\cdot \bm{a_2}[\sigma_z]_{\sigma\sigma}/2}\tilde{c}_{\sigma}^{\dagger}(\bm r)\tilde{c}_{\sigma}(\bm r-\bm{a}_2) \nonumber\\
~~~~~~ &+&e^{-i\bm{K}\cdot \bm{a_3}[\sigma_z]_{\sigma\sigma}/2}\tilde{c}_{\sigma}^{\dagger}(\bm r)\tilde{c}_{\sigma}(\bm r+\bm{a}_3)+h.c.]\nonumber\\
&=&\sum_{\bm{r},\sigma}e^{-i2\pi/3[\sigma_z]_{\sigma\sigma}}[\tilde{c}_{\sigma}^{\dagger}(\bm r)\tilde{c}_{\sigma}(\bm r+\bm{a}_1)+\tilde{c}_{\sigma}^{\dagger}(\bm r)\tilde{c}_{\sigma}(\bm r-\bm{a}_2)+\tilde{c}_{\sigma}^{\dagger}(\bm r)\tilde{c}_{\sigma}(\bm r+\bm{a}_3)+h.c.],
\eeqn
where $\bm{a}_3=\bm{a}_2-\bm{a}_1$ and we have defined $\tilde{c}_{\sigma}(\bm r)=\sum_{\sigma^{\prime}}[\mathcal{U}(\bm r)]_{\sigma\sigma^{\prime}}c_{\sigma^{\prime}}(\bm r)$. Therefore, $\tilde{H}_1$ describes a system defined on a triangular lattice with complex hopping $te^{-i2\pi/3\sigma_z}$ under uniform Zeeman field along the $x$-direction. After the Fourier transformation,  $\tilde{H}_1$ can be written as 
\beqn
\tilde{H}_1(\bm k)=J\sigma_x+\tilde{f}_0(\bm k)\sigma_0+\tilde{f}_z(\bm k)\sigma_z,
\eeqn
where $\tilde{f}_0(\bm k)=-t\cos k_xa-2t\cos k_xa/2\cos \sqrt{3}k_ya/2$ and $\tilde{f}_z(\bm k)=\sqrt{3}t(\sin k_xa-2\sin k_xa/2\cos \sqrt{3}k_ya/2)$. 
It is noted that the local transformation $\mathcal{U}(\bm r)$ does not change the spin expectation value $s_z(\bm k)$ due to $[U(\bm r), \sigma_z]=0$. Therefore, we have $s_z(\bm k)=\tilde{s}_z(\bm k)$,  where  $\tilde{s}_z(\bm k)=\langle \tilde{\psi}_{\bm{k}} | \sigma_z | \tilde{\psi}_{\bm{k}} \rangle$ with $| \tilde{\psi}_{\bm{k}} \rangle$ being the eigenstate of $\tilde{H}_1(\bm k)$. It can be directly derived that $\tilde{s}_z(\bm k)=\tilde{f}_z(\bm k)/\sqrt{J^2+\tilde{f}_z^2(\bm k)}$.  By expanding $\tilde{f}_z(\bm k)$ near the $\Gamma$ point, we have $\tilde{f}_z(\bm k)\approx k_x(k_x^2-3k_y^2)$, and therefore 
$s_z(\bm k)\propto k_x(k_x^2-3k_y^2)$. From $\tilde{H}_1$, we can see that the effective SOC determined by $\tilde{f}_z(\bm k)$ is proportional to $t$, which can be substantial in realistic systems. In addition to SOC term, the term $J\sigma_x$ describe an effective Zeeman field, which lift the Kramers degeneracy at time-reversal invariant momenta. Therefore, the band structure of $H_1$ is described by the interplay between effective SOC and Zeeman field, which can be a general characteristic of OPMs.

\subsection{Theoretical model for type-II OPMs}
In the main text, we construct a theoretical model denoted by $H_2$ to realize type-II OPMs. $H_2$ is defined on a square lattice and comprises sixteen magnetic sublattices per unit cell. The Hamiltonian of $H_2$ is given by
\beqn
H_{2} = \sum_{\langle ij\alpha\beta\rangle,\sigma} t c_{i\alpha\sigma}^\dagger c_{j\beta\sigma} 
+ J \sum_{i,\alpha,\sigma,\sigma^{\prime}} \bm{m}_{\alpha} \cdot c_{i\alpha\sigma}^\dagger \bm{\sigma}_{\sigma\sigma'} c_{i\alpha\sigma'},
\eeqn
where $i$ and $j$ label unit cells, $\alpha=1,\cdots,16$ indexes the sublattices, and $\langle ij\alpha\beta\rangle$ denotes summation over nearest-neighbor sites. The magnetic moment directions for each sublattice are
\beqn
&&\bm{m}_{1}=(1,1,0),\quad \bm{m}_{2}=(1,-1,0),\quad \bm{m}_{3}=(-1,-1,0),\quad \bm{m}_{4}=(-1,1,0),\nonumber\\
&&\bm{m}_{5}=(0,1,1),\quad \bm{m}_{6}=(0,-1,1), \quad \bm{m}_{7}=(0,-1,-1),\quad \bm{m}_{8}=(0,1,-1),\nonumber\\
&&\bm{m}_{9}=(0,-1,-1),\quad \bm{m}_{10}=(0,1,-1),\quad \bm{m}_{11}=(0,1,1),\quad \bm{m}_{12}=(0,-1,1),\nonumber\\
&&\bm{m}_{13}=(1,-1,0),\quad \bm{m}_{14}=(1,1,0),\quad \bm{m}_{15}=(-1,1,0),\quad \bm{m}_{16}=(-1,-1,0).
\eeqn
The arrangement of these magnetic moments is shown in Figure.~1(c). $H_2$ respects the symmetries $g_3 = \{T||I|\tau\}$ and $g_4 = \{U_z(\pi)||P\}$ and therefore belongs to type-II OPMs with nonzero spin textures $s_{x,y}(\bm{k})$. Additionally, $H_2$ respects the symmetries $\hat{M}_y=\{U_x(\pi)||M_y\}$ and $\hat{M}_x=\{U_y(\pi)||M_x\}$. These spin group symmetries generate the point group $\tilde{G}=C_{2v}$, which is generated by $R_{2z}$ and $M_x$. In particular, $s_x(\bm{k})$ and $s_y(\bm{k})$ form $B_1$ and $B_2$ representations of $C_{2v}$, respectively, which are associated with the symmetry constraints
\beqn
&& s_{x}(k_x,k_y)=-s_{x}(-k_x,k_y),\nonumber\\ 
&& s_{y}(k_x,-k_y)=-s_{y}(k_x,-k_y),\nonumber\\
&& s_{x,y}(k_x,k_y)=s_{x,y}(-k_x,-k_y).
\eeqn
The lowest-order basis functions of $B_1$ and $B_2$ representations are $x$ and $y$, respectively. Consequently, the NSS of $H_2$ can be described by $k_x \sigma_x + k_y \sigma_y$, where the superposition coefficients are omitted.

%along with an additional symmetry $g_r = \{I || R_{3z}\}$. These symmetries enforce $s_z(\bm{k}) = -s_z(-\bm{k}) = s_z(R_{3z} \bm{k})$, leading to $s_z(R_{6z} \bm{k}) = -s_z(\bm{k})$. The lowest-order term compatible with these constraints takes the form $s_z(\bm k)\approx k_x(k_x^2-3k_y^2)$, consistent with analytical derivations \cite{supp}. This results in a $f$-wave pattern, as depicted in Fig.~\ref{Fig1}(b). The spin texture $s_z(\bm k)$ forms the $B_1$ representation of point group $6mm$ produced by the spin group symmetries.

%$\tilde{g}_1=\{U_x(\pi)||M_y\}$, and $\tilde{g}_2=\{U_y(\pi)||M_x\}$. 

%is constructed to capture the key symmetries of CeIn$_3$N. The arrangement of local magnetic moments in a unit cell is illustrated in Fig.~\ref{Fig1}(c). $H_2$ respects the symmetries $g_3=\{T||I|\tau\}$, $g_4=\{U_z(\pi)||P\}$, $\tilde{g}_1=\{U_x(\pi)||M_y\}$, and $\tilde{g}_2=\{U_y(\pi)||M_x\}$, which are identical to those of CeIn$_3$N. These symmetries impose the constraints $s_x(-k_x,k_y)=-s_x(k_x,k_y)$ and $s_y(k_x,-k_y)=-s_y(k_x,k_y)$, in agreement with the numerical results [Fig.~\ref{Fig1}(d)]. Consequently, the NSS of $H_2$ can be approximatively described by $a_{100}^{x}k_x\sigma_x+a_{010}^{y}k_y\sigma_y$ \cite{Yang2024}, exhibiting a $p$-wave pattern.

%, which exhibits a sign pattern proportional to $k_x(k_x^2-3k_y^2)$, similar to the spin expectation value $s_z(\bm k)$ of Hamiltonian $H_2$.

\subsection{Theoretical models for type-III OPMs}
 The model Hamiltonian of the type-III OPMs defined on a bilayer breathing kagome lattice in the main text is:
\beqn
H_3&=  \sum\limits_{ \alpha\beta\sigma\ell}(t_1\sum\limits_{\langle ij\rangle}c_{i\alpha\sigma\ell}^{\dagger}c_{j\beta\sigma\ell}+ t_2 \sum\limits_{\bar{\langle ij\rangle}}c_{i\alpha\sigma\ell}^{\dagger}c_{j\beta\sigma\ell})\nonumber\\
&+ \sum\limits_{i\alpha\sigma\ell\ell^{\prime}}t_{\perp}c_{i\alpha\sigma\ell}^{\dagger}c_{i\alpha\sigma\ell^{\prime}}+\sum\limits_{i\alpha\sigma\ell}J\bm{m}_{\alpha\ell}\cdot c_{i\alpha\sigma\ell}^{\dagger} \bm{\sigma} c_{i\alpha\sigma\ell},
\eeqn
where $\alpha, \beta = \text{A, B, C}$ denote the three sublattices, and $\ell, \ell' = \mathfrak{b}, \mathfrak{t}$ represent the bottom and top layers, respectively. 
The first and second terms describe intracell and intercell hopping with amplitudes $t_1$ and $t_2$, respectively. The third term represents interlayer hopping with amplitude $t_\perp$ and the fourth term accounts for noncoplanar magnetic orders described by $\bm{m}_{\alpha\ell}$. We set $t_1\neq t_2$ and $\bm{m}_{\alpha\mathfrak{b}} = -\bm{m}_{\alpha\mathfrak{t}}$. $\bm{m}_{\alpha\ell}$ has a canting angle $\theta=\pi/3$ and in-plane components form an all-in–all-out 
 structure, namely, $\bm{m}_{\text{A}\mathfrak{t}} = (1/4, \sqrt{3}/4, \sqrt{3}/2)$, $\bm{m}_{\text{B}\mathfrak{t}} = (-1/4, \sqrt{3}/4, \sqrt{3}/2)$, and $\bm{m}_{\text{C}\mathfrak{t}} = (0, 1/2, \sqrt{3}/2)$. We take the model parameters as $t_{\perp}=t_1=1$ and $t_2=J=4$. In Figure.1~(d), $\mu=-3.3$.
 
$H_3$ respects the symmetry $\tilde{T} = \{T || M_z\}$ and therefore belongs to type-III OPMs with nonzero spin expectation values $s_{x,y,z}(\bm k)$. Additionally, $H_3$ preserves the symmetries $\hat{C}_{2y}=\{U_x(\pi)||R_{2y}\}$ and $\hat{C}_{3z}=\{U_z(2\pi/3)||R_{3z}\}$. These spin group symmetries generate the point group $\tilde{G}=C_{6v}$. According to Eq.~\eqref{seq1}, we obtain the following symmetry constraints:
\beqn
&&s_z(R_{6z} \bm{k})=-s_z(\bm k),\quad s_z(k_x,k_y)=-s_z(-k_x,k_y),\nonumber\\
&&s_x(R_{6z}\bm k)=s_x(\bm k)/2-\sqrt{3}s_y(\bm k)/2,\quad s_y(R_{6z}\bm k)=\sqrt{3}s_x(\bm k)/2+s_y(\bm k)/2,\nonumber\\
&&s_x(M_x\bm k)=s_x(\bm k),\quad s_y(M_y\bm k)=-s_y(\bm k).
\eeqn
Consequently, $s_z(\bm k)$ forms a $B_1$ representation of $C_{6v}$ with basis function $x(x^2-3y^2)$. Meanwhile, the spin textures $s_x(\bm k)$ and $s_y(\bm k)$ form a two-dimensional $E_1$ representation of $C_{6v}$ with basis functions $(y,x)$. Thus, near the $\Gamma$ point, the NSS of $H_3$ can be described by $k_y\sigma_x+k_x\sigma_y+k_x(k_x^2-3k_y^2)\sigma_z$.

\section{Candidate materials for odd-parity magnets}
In this section, we detail the procedure for identifying candidate materials for OPMs. In Ref.~\onlinecite{XiaoZhenyu2024}, the authors utilized symmetries that leave generic $\bm k$ points invariant to screen the magndate database for materials exhibiting NSS. The candidate materials are classified based on the number of nonzero components ($d_s = 1$, $2$, and $3$, respectively) of $\bm S(\bm{k})$. We search for type-I (\(d_s=1\)), type-II (\(d_s=2\)), and type-III (\(d_s=3\)) OPMs among these materials.

Additionally, \citet{CheXiaobing2024} developed an online tool named FindSpinGroup, which can identifies all spin space group operations for materials in the magndata database. Using this tool, we analyzed the candidate materials with NSS identified in \citet{XiaoZhenyu2024} to determine the target spin symmetries relevant to OPMs.  Our analysis reveals 33 candidate materials for OPMs, as summarized in Table~\ref{tab3}, including 25 type-I, 1 type-II, and 7 type-III candidates. We emphasize that in collinear magnetic orders, the emergence of OPMs requires breaking \(C_2{T}\) symmetry through complex hopping of electrons, a condition incompatible with the framework established in Refs.~\cite{XiaoZhenyu2024,CheXiaobing2024}. Consequently, all identified OPM candidates exhibit coplanar or non-coplanar magnetic orders.

For a given material with spin space group \(G\), the corresponding point group \(\tilde{G} = \{s_g R_g \mid g \in G\}\) can be derived. The spin textures form a \(d_s\)-dimensional representation of \(\tilde{G}\). The lowest-order basis functions of this representation determine the functional form of $\bm S(\bm k)$ near the \(\Gamma\) point. In Tables~\ref{stab1} and \ref{stab2}, we list the spin space group \(G\), the symmetries \(g_{1-5}\) (where present), and the nonzero spin texture components for each candidate material. In Tables~\ref{stab3} and \ref{stab4}, we summarize the point group \(\tilde{G}\), the representation formed by the spin textures under \(\tilde{G}\), the basis functions of this representation, the functional form of the spin textures near the \(\Gamma\) point, the term \(\bm S(\bm k) \cdot \bm{\sigma}\) describing the NSS, and the partial wave channels associated with the NSS.

\begin{table*}
\centering
\setlength\tabcolsep{3pt}
\renewcommand{\arraystretch}{2.1}
\caption{Spin space group, the existing symmetries of $g_{1-5}$, and nonzero components of spin textures for each candidate material. }
\begin{tabular}{|c|c|c|c|c|}
\hline
Types & Materials & SSG &  symmetries $g_{1-5}$ & nonzero components\\
\hline
\multirow{14}{*}{\makecell{type-I\\ OPMs}}& Sr$_2$Fe$_3$Se$_2$O$_3$ & $ C^1 m^{2_{010}}(1/2~0 ~ 1/2)^{m_{001}}(1/2 ~ 0 ~ 0)$ & $g_{1}=\{2_{010}||\tau\}$, $g_3=\{-1||1|\tau\}$ & $s_y(\bm k)=-s_y(-\bm k)$\\
\cline{2-5}
~&CsFeCl$_3$&$ P^{2_{100}} 6_3 / ^1m^{2_{100}}m^1c|(3^1_{001}, 3^1_{001}, 1)^m 1$& $g_1=\{3_{001}^{1}||1|\tau\}$,  $g_5=\{m_{001}||1\}$ & $s_z(\bm k)=-s_z(-\bm k)$\\
\cline{2-5}
~& ThMn$_2$ & $ P^{2_{\frac{5\pi}{6}}}6_3 /^1m^{2_{\frac{\pi}{3}}}m^{2_{001}}c|(3^1_{001}, 3^1_{001}, 1)^m 1$ & $g_1=\{3_{001}||1|\tau\}$, $g_5=\{m_{001}||1\}$ & $s_z(\bm k)=-s_z(-\bm k)$ \\
\cline{2-5}
~&EuIn$_2$As$_2$& $P^{6_{001}^1}6_3 / ^{2_{100}}m^1m{^{6_{001}^1}}c(1, 1, 3^1_{001})^m1$ & $g_1=\{3_{001}^1||1|\tau\}$, $g_5=\{m_{001}||1\}$ & $s_z(\bm k)=-s_z(-\bm k)$\\
\cline{2-5}
~&CsMnBr$_3$& $ P^{2_{\frac{\pi}{3}}}6_3 /^{2_{001}}m^{2_{\frac{5\pi}{6}}}m^{2_{001}}c|(3^1_{001}, 3^1_{001}, 1)^m 1$ & $g_1=\{3_{001}||1|\tau\}$, $g_5=\{m_{001}||1\}$  & $s_z(\bm k)=-s_z(-\bm k)$\\
\cline{2-5}
~&RbFeCl$_3$ &$ P^{2_{\frac{\pi}{3}}}6_3 /^1m^{2_{\frac{\pi}{3}}}m^1c|(3^1_{001}, 3^1_{001}, 1)^m 1$& $g_1=\{3_{001}||1|\tau\}$, $g_5=\{m_{001}||1\}$  & $s_z(\bm k)=-s_z(-\bm k)$\\
\cline{2-5}
~& RbNiCl$_3$& $ P^{2_{010}}6_3 /^{2_{100}}m^{2_{100}}c^{2_{001}}m^m 1$ & $g_1=\{3_{001}||1|\tau\}$, $g_5=\{m_{001}||1\}$ & $s_z(\bm k)=-s_z(-\bm k)$\\
\cline{2-5}
~&Ba$_3$CoSb$_2$O$_9$ & $P^{2_{100}}6_3 /^{2_{001}}m^{2_{010}} m^{2_{001}}c|(3^2_{001}, 3^2_{001}, 1)^m 1$ & $g_1=\{3_{010}||1|\tau\}$, $g_5=\{m_{010}||1\}$ & $s_y(\bm k)=-s_y(-\bm k)$\\
\cline{2-5}
~&CsFe(MoO$_4$)$_2$& $ P^{2_{\frac{\pi}{6}}}-3|(3^1_{001}, 3^1_{001}, 2_{001})^m 1 $&$g_1=\{3_{001}||1|\tau\}$, $g_5=\{m_{001}||1\}$ & $s_z(\bm k)=-s_z(-\bm k)$\\
\cline{2-5}
~&Er$_2$Pt & $P^1m^{2_{001}}n^{2_{001}}2_1|(1, 2_{001}, 1)^m 1$ & $g_3=\{-1||1|\tau\}$, $g_5=\{m_{100}||1\}$ & $s_x(\bm k)=-s_x(-\bm k)$\\
\cline{2-5}
~&DyBe$_{13}$ & $I^14/^{2_{010}}m^{2_{001}}c^{2_{001}}m|(1,1,1; 2_{001})^m 1$ & $g_3=\{-1||1|\tau\}$, $g_5=\{m_{001}||1\}$ & $s_z(\bm k)=-s_z(-\bm k)$\\
\cline{2-5}
~& TbC$_2$ & $P^1m^{1}m^{2_{010}}m|(1, 1, 2_{001})^m 1$ &$g_3=\{-1||1|\tau\}$, $g_5=\{m_{001}||1\}$ & $s_z(\bm k)=-s_z(-\bm k)$\\
\cline{2-5}
~& Ca$_2$Cr$_2$O$_5$ & $P^1m^{1}a^12^{2_{001}}(1/2~1/2~1/2)^m 1$ & $g_3=\{-1||1|\tau\}$, $g_5=\{m_{001}||1\}$ & $s_z(\bm k)=-s_z(-\bm k)$\\
\cline{2-5}
~& Tm$_5$Pt$_2$In$_4$ & $C^1m^{2_{001}}(1/2~0~0)^m 1$ & $g_3=\{-1||1|\tau\}$, $g_5=\{m_{010}||1\}$ & $s_y(\bm k)=-s_y(-\bm k)$\\ 
\cline{2-5}
~&La$_{1/3}$Ca$_{2/3}$MnO$_3$& $P^{2_{100}}m^{2_{001}}c^{2_{010}}2_1|(1, 2_{100}, 1)^m 1$ &$g_3=\{-1||1|\tau\}$, $g_5=\{m_{100}||1\}$  & $s_x(\bm k)=-s_x(-\bm k)$\\
\cline{2-5}
~&La$_{3/8}$Ca$_{5/8}$MnO$_3$& $P^{2_{100}}m^{2_{001}}c^{2_{010}}2_1|(1, 2_{100}, 1)^m 1$ &$g_3=\{-1||1|\tau\}$, $g_5=\{m_{100}||1\}$  & $s_x(\bm k)=-s_x(-\bm k)$\\
\cline{2-5}
~& KFe(PO$_3$F)$_2$&  $P^{2_{\frac{11\pi}{12}}}-3|(3_{001}^1, 3_{001}^1, 4_{001}^1)^m 1$ &$g_3=\{-1||1|\tau\}$, $g_5=\{m_{001}||1\}$  & $s_z(\bm k)=-s_z(-\bm k)$\\
\cline{2-5}
~& NiCr$_2$O$_4$& $I^{2_{100}}2_{1}^{~2_{001}}2_1|(1,1,1;2_{010})^m 1$ & $g_3=\{-1||1|\tau\}$, $g_5=\{m_{010}||1\}$ & $s_y(\bm k)=-s_y(-\bm k)$\\
\cline{2-5}
~&PrMn$_2$O$_5$ & $P^1m^{2_{010}}c^{2_{010}}2_1(1,2_{100},1)^m 1$  & $g_3=\{-1||1|\tau\}$, $g_5=\{m_{100}||1\}$ & $s_x(\bm k)=-s_x(-\bm k)$\\
\cline{2-5}
~&GdMn$_2$O$_5$ & $P^1m^{2_{001}}c^{2_{001}}2_1|(1,2_{100},1)^m 1$ & $g_3=\{-1||1|\tau\}$, $g_5=\{m_{100}||1\}$ & $s_x(\bm k)=-s_x(-\bm k)$\\
\cline{2-5}
~& Sr$_2$FeO$_3$Cl &  $P^{2_{010}}4/^{2_{110}}n^{2_{010}}m^1m|(2_{001},2_{001},1)^m 1$&  $g_3=\{-1||1|\tau\}$, $g_5=\{m_{001}||1\}$ & $s_z(\bm k)=-s_z(-\bm k)$\\
\hline
\end{tabular}
\label{stab1}
\end{table*}

\begin{table*}
\centering
\setlength\tabcolsep{3pt}
\renewcommand{\arraystretch}{2.1}
\caption{Spin space group, the existing symmetries of $g_{1-5}$, and nonzero components of spin textures for each candidate material.}
\begin{tabular}{|c|c|c|c|c|}
\hline
Types & Materials & SSG &  symmetries $g_{1-5}$ & nonzero components\\
\hline
\cline{2-5}
\multirow{5}{*}{\makecell{type-I\\ OPMs}}& CoNb$_2$O$_6$ & $P^{2_{010}}2_1^{~2_{010}}2_1^{~1}2|(1,1,2_{001})^m 1$ & $g_3=\{-1||1|\tau\}$, $g_5=\{m_{001}||1\}$ & $s_z(\bm k)=-s_z(-\bm k)$\\
\cline{2-5}
~&CeNiAsO & $P^{2_{100}}2_1/^1m|(2_{001},1,1)^m 1$& $g_3=\{-1||1|\tau\}$, $g_5=\{m_{001}||1\}$ & $s_z(\bm k)=-s_z(-\bm k)$\\
\cline{2-5}
~& DyMn$_2$O$_5$ & $P^{1}m^{2_{001}}c^{2_{001}}2_1|(1,2_{100},1)^m 1$ & $g_3=\{-1||1|\tau\}$, $g_5=\{m_{100}||1\}$ & $s_x(\bm k)=-s_x(-\bm k)$\\
\cline{2-5}
~& BiMn$_2$O$_5$ & $ P^{2_{100}} m^{2_{010}}c^{2_{001}}2_1|(2_{100},2_{100}, 1)^{m}1$ & $g_3=\{-1||1|\tau\}$, $g_5=\{m_{100}||1\}$ & $s_x(\bm k)=-s_x(-\bm k)$\\
\hline
\multirow{1}{*}{\makecell{type-II\\ OPMs}}& Ce$_3$InN & $ P^{-4_{001}^3} 4/^1m^{2_{010}}m^{m_{110}}m|(-1,-1,1)$ & $g_3=\{-1||1|\tau\}$, $g_4=\{2_{001}||-1\}$ & $s_{x,y}(\bm k)=-s_{x,y}(-\bm k)$\\
\hline
\multirow{8}{*}{\makecell{type-III\\ OPMs}}& MgV$_2$O$_4$ & $ F^{m_{100}} 2^{2_{010}}2^{m_{001}}2|(1,1,1;-1,-1,1)$ & $g_3=\{-1||1|\tau\}$& $s_{x,y,z}(\bm k)=-s_{x,y,z}(-\bm k)$\\
\cline{2-5}
~& Mn$_5$Si$_3$ & $ P^{1} m^1c^12_1^{-1}(0~1/2~1/2)$ & $g_3=\{-1||1|\tau\}$& $s_{x,y,z}(\bm k)=-s_{x,y,z}(-\bm k)$\\
\cline{2-5}
~&Ba(TiO)Cu$_4$(PO$_4$)$_4$ & $P^{-4_{001}} 4^{m_{100}}2_1^{~2_{110}}2|(1,1,-1)$ & $g_3=\{-1||1|\tau\}$& $s_{x,y,z}(\bm k)=-s_{x,y,z}(-\bm k)$\\
\cline{2-5}
~&Dy$_2$Co$_3$Al$_9$ & $A^{m_{100}} m^{2_{010}}m^{m_{001}}2|(-1,1,1;1)$ & $g_3=\{-1||1|\tau\}$& $s_{x,y,z}(\bm k)=-s_{x,y,z}(-\bm k)$\\
\cline{2-5}
~&DyFeWO$_6$ & $P^{m_{001}} c|(-1,-1,1)$& $g_3=\{-1||1|\tau\}$& $s_{x,y,z}(\bm k)=-s_{x,y,z}(-\bm k)$\\
\cline{2-5}
~&Ho$_2$Cu$_2$O$_5$ & $P^{m_{001}} 2_1|(-1,1,1)$&$g_3=\{-1||1|\tau\}$& $s_{x,y,z}(\bm k)=-s_{x,y,z}(-\bm k)$\\
\cline{2-5}
~&BaFe$_2$Se$_3$ & $P^{2_{001}} m|(-1,-1,-1)$& $g_3=\{-1||1|\tau\}$& $s_{x,y,z}(\bm k)=-s_{x,y,z}(-\bm k)$\\
\hline
\end{tabular}
\label{stab2}
\end{table*}

\begin{table*}
\centering
\setlength\tabcolsep{5pt}
\renewcommand{\arraystretch}{2.1}
\caption{The emergent point group $\tilde{G}$, the formed representation by $\bm S(\bm k)$ of $\tilde{G}$, the basis function of the formed representation, the functional form of spin texture near the $\Gamma$ point, the term $\bm {S}(\bm k)\cdot \bm{\sigma}$, and the partial wave channel of NSS for each candidate materials. }
\begin{tabular}{|c|c|c|c|c|c|c|c|c|}
\hline
Types&Materials &PG ($\tilde{G}$)&  Rep of $\bm S(\bm k)$& basis function & $\bm S(\bm k)$ & $\bm S(\bm k)\cdot \bm{\sigma}$ & channel\\
\hline
\multirow{14}{*}{\makecell{type-I\\ OPMs}}& Sr$_2$Fe$_3$Se$_2$O$_3$ & $2/m$ & $B_u$ & $x,z$ & $s_y(\bm k)\approx k_x+k_z$ &$(k_x+k_z)\sigma_y$ & $p$-wave\\
\cline{2-8}
~&CsFeCl$_3$& $6/mmm$ & $B_{2u}$ & $y(y^2-3x^2)$ & $s_z(\bm k)\approx k_y(k_y^2-3k_x^2)$ &$k_y(k_y^2-3k_x^2) \sigma_z$ & $f$-wave\\
\cline{2-8}
~& ThMn$_2$ & $6/mmm$ & $B_{2u}$ & $y(y^2-3x^2)$ & $s_z(\bm k)\approx k_y(k_y^2-3k_x^2)$ & $k_y(k_y^2-3k_x^2)\sigma_z$ & $f$-wave\\
\cline{2-8}
~&EuIn$_2$As$_2$&  $6/mmm$ & $A_{2u}$& $z$&  $s_z(\bm k)\approx k_z$& $k_z\sigma_z$ & $p$-wave\\
\cline{2-8}
~&CsMnBr$_3$ & $6/mmm$ & $B_{2u}$ & $y(y^2-3x^2)$ & $s_z(\bm k)\approx k_y(k_y^2-3k_x^2)$ & $k_y(k_y^2-3k_x^2)\sigma_z$ & $f$-wave\\
\cline{2-8}
~&RbFeCl$_3$ & $6/mmm$ & $B_{2u}$ & $y(y^2-3x^2)$ & $s_z(\bm k)\approx k_y(k_y^2-3k_x^2)$ & $k_y(k_y^2-3k_x^2)\sigma_z$ & $f$-wave \\
\cline{2-8}
~&Ba$_3$CoSb$_2$O$_9$ & $6/mmm$ & $B_{2u}$ & $y(y^2-3x^2)$ & $s_y(\bm k)\approx k_y(k_y^2-3k_x^2)$ & $k_y(k_y^2-3k_x^2)\sigma_y$ & $f$-wave \\
\cline{2-8}
~&CsFe(MoO$_4$)$_2$& $\bar{3}$ &$A_{u}$ & $z$ & $s_z(\bm k)\approx k_z$ & $k_z\sigma_z$ & $p$-wave\\
\cline{2-8}
~& Er$_2$Pt & $mmm$ & $B_{2u}$ & $y$& $s_x(\bm k)\approx k_y$ & $k_y\sigma_x$ & $p$-wave\\
\cline{2-8}
~&DyBe$_{13}$ &4/mmm & $A_{2u}$ & $z$ & $s_z(\bm k)\approx k_z$ & $k_z\sigma_z$ & $p$-wave\\
\cline{2-8}
~&TbC$_{2}$ & $mmm$ & $B_{1u}$ & $z$ & $s_z(\bm k)\approx k_z$ & $k_z\sigma_z$ & $p$-wave \\
\cline{2-8}
~&Ca$_2$Cr$_2$O$_5$ & $mmm$ & $B_{1u}$ & $z$ & $s_z(\bm k)\approx k_z$ & $k_z\sigma_z$ & $p$-wave  \\
\cline{2-8}
~& Tm$_5$Pt$_2$In$_4$& $2/m$ & $B_{u}$  &$x,z$ & $s_y(\bm k)\approx k_x+k_z$  & $(k_x+k_z)\sigma_y$ & $p$-wave\\
\cline{2-8}
~& La$_{1/3}$Ca$_{2/3}$MnO$_3$ & $mmm$ & $B_{2u}$ & $y$ &  $s_x(\bm k)\approx k_y$  & $k_y\sigma_x$ & $p$-wave\\
\cline{2-8}
~& La$_{3/8}$Ca$_{5/8}$MnO$_3$ & $mmm$ & $B_{2u}$ & $y$ & $s_x(\bm k)\approx k_y$  & $k_y\sigma_x$ & $p$-wave\\
\cline{2-8}
~& KFe(PO$_3$F)$_2$ & $\bar{3}$ & $A_u$ & $z$ & $s_z(\bm k)\approx k_z$  & $k_z\sigma_z$ & $p$-wave \\
\cline{2-8}
~& NiCr$_2$O$_4$& $D_{2h}$ & $B_{2u}$ & $y$ & $s_y(\bm k)\approx k_y$  & $k_y\sigma_y$ & $p$-wave\\
\cline{2-8}
~&PrMn$_2$O$_5$ & $mmm$ & $B_{2u}$ & $y$ & $s_x(\bm k)\approx k_y$  & $k_y\sigma_x$ & $p$-wave\\
\cline{2-8}
~&GdMn$_2$O$_5$ & $mmm$ & $B_{2u}$ & $y$ & $s_x(\bm k)\approx k_y$  & $k_y\sigma_x$ & $p$-wave \\
\cline{2-8}
~ &Sr$_2$FeO$_3$Cl & $D_{4h}$ & $B_{1u}$ & $xyz$ & $s_z(\bm k)\approx k_xk_yk_z$  & $k_xk_yk_z\sigma_z$ & $p$-wave \\
\hline
\end{tabular}
\label{stab3}
\end{table*}

\begin{table*}
\centering
\setlength\tabcolsep{3pt}
\renewcommand{\arraystretch}{2.1}
\caption{The emergent point group $\tilde{G}$, the formed representation by $\bm S(\bm k)$ of $\tilde{G}$, the basis function of the formed representation, the functional form of spin texture near the $\Gamma$ point, the term $\bm {S}(\bm k)\cdot \bm{\sigma}$, and the partial wave channel of NSS for each candidate materials. }
\begin{tabular}{|c|c|c|c|c|c|c|c|}
\hline
Types&Materials &PG ($\tilde{G}$)&  Rep of $\bm S(\bm k)$& basis function &  $\bm S(\bm k)$ & $\bm S(\bm k)\cdot \bm{\sigma}$ & channel\\
\hline
\multirow{3}{*}{\makecell{type-I\\ OPMs}}~ &CoNb$_2$O$_6$ & $mmm$ & $B_{1u}$ & $z$  & $s_z(\bm k)\approx k_z$  & $k_z\sigma_z$ & $p$-wave\\
\cline{2-8}
~& CeNiAsO & $2/m$ & $B_u$ & $x,z$ & $s_z(\bm k)\approx k_x+k_z$ &$(k_x+k_z)\sigma_z$ & $p$-wave\\
\cline{2-8}
~& DyMn$_2$O$_5$& $mmm$ & $B_{2u}$ & $y$ &  $s_x(\bm k)\approx k_y$ &$k_y\sigma_x$  & $p$-wave \\
\cline{2-8}
~& CeNiAsO & $2/m$ & $B_u$ & $x,z$ & $s_z(\bm k)\approx k_x+k_z$ & $(k_x+k_z)\sigma_z$ & $p$-wave\\
\cline{2-8}
~& BiMn$_2$O$_5$ & $mmm$ & $B_{2u}$ & $y$ & $s_y(\bm k)\approx k_y$ & $k_y\sigma_x$ & $p$-wave\\
\hline
\makecell{type-II\\ OPMs}& Ce$_3$InN & $4/mmm$ & $E_u$ & $(x,y)$ & $\makecell{s_x(\bm k)\approx k_x\\ s_y(\bm k)\approx k_y}$  & $(k_x\sigma_x+k_y\sigma_y)$  & $p$-wave\\
\hline
\multirow{7}{*}{\makecell{type-III\\ OPMs}}& MgV$_2$O$_4$ &  $mmm$ & $B_{3u} \bigoplus B_{2u}  \bigoplus B_{1u}$ & $(x,y,z)$ & $\makecell{s_x(\bm k)\approx k_x\\ s_y(\bm k)\approx k_y\\ s_z(\bm k)\approx k_z} $ & $\makecell{k_x\sigma_x+\\ k_y\sigma_y+\\ k_z\sigma_z}$ & $p$-wave\\
\cline{2-8}
~& Mn$_5$Si$_3$ & $mmm$ & $B_{1u}\bigoplus B_{1u} \bigoplus B_{1u}$ & $(z,z,z)$ & $s_{x,y,z}(\bm k)\approx k_z $& $\makecell{k_z\sigma_x+\\ k_z\sigma_y+\\ k_z\sigma_z}$ & $p$-wave\\
\cline{2-8}
~& Ba(TiO)Cu$_4$(PO$_4$)$_4$ & $4/mmm$ & $E_u \bigoplus A_{2u}$ & $(x,y)\bigoplus (z) $  & $\makecell{s_x(\bm k)\approx k_x\\ s_y(\bm k)\approx k_y\\ s_z(\bm k)\approx k_z} $ &  $\makecell{k_x\sigma_x+\\ k_x\sigma_y+\\k_z\sigma_z}$ & $p$-wave \\
\cline{2-8}
~& Dy$_2$Co$_3$Al$_9$ & $mmm$ & $B_{2u} \bigoplus B_{3u}  \bigoplus A_{u}$ & $(y,x,xyz)$ & $\makecell{s_x(\bm k)\approx k_y\\ s_y(\bm k)\approx k_x\\ s_z(\bm k)\approx k_xk_yk_z} $& $\makecell{k_y\sigma_x+\\
k_x\sigma_y+\\k_xk_yk_z\sigma_z}$ & $\makecell{s_{x,y}(\bm k), p-\text{wave}\\ s_{z}(\bm k), f-\text{wave}}$\\
\cline{2-8}
~&  Ho$_2$Cu$_2$O$_5$ & $2/m$ & $B_{u} \bigoplus A_{u}  \bigoplus B_{u}$ & $(x,z) \bigoplus y \bigoplus (x,z)$ & $\makecell{s_x(\bm k)\approx k_x+k_z\\ s_y(\bm k)\approx k_y\\ s_z(\bm k)\approx k_x+k_z} $& $\makecell{(k_x+k_z)\sigma_x\\ +k_y\sigma_y+\\(k_x+k_z)\sigma_z}$  & $p$-wave\\
\cline{2-8}
~& BaFe$_3$Se$_3$ & $2/m$ & $A_{u} \bigoplus B_{u}  \bigoplus A_{u}$ & $y \bigoplus (x,z) \bigoplus y$ & $\makecell{s_x(\bm k)\approx k_y\\ s_y(\bm k)\approx k_y+k_z\\ s_z(\bm k)\approx k_y} $&  $\makecell{ k_y\sigma_x+\\(k_x+k_z)\sigma_y\\+k_y\sigma_z}$  & $p$-wave\\
\cline{2-8}
~&TbSbTe & $2/m$ & $B_{u} \bigoplus A_{u}  \bigoplus B_{u}$ & $(x,z) \bigoplus y \bigoplus (x,z)$ &  $\makecell{s_x(\bm k)\approx k_x+k_z\\ s_y(\bm k)\approx k_y\\ s_z(\bm k)\approx k_x+k_z} $ & $\makecell{(k_x+k_z)\sigma_x\\ +k_y\sigma_y+\\(k_x+k_z)\sigma_z}$ & $p$-wave\\
\hline
\end{tabular}
\label{stab4}
\end{table*}

\end{widetext}

\end{document}